# Self-Organized Criticality in Proteins:
# Hydropathic Roughening Profiles of G-Protein Coupled Receptors

**J. C. Phillips**

Dept. of Physics and Astronomy, Rutgers University, Piscataway, N. J., 08854

Abstract

Proteins appear to be the most dramatic natural example of self-organized criticality (SOC), a concept that explains many otherwise apparently unlikely phenomena. Protein conformational functionality is often dominated by long-range hydro(phobic/philic) interactions which both drive protein compaction and mediate protein-protein interactions. Superfamily transmembrane GPCR are the largest family of proteins in the human genome; their amino acid sequences form the largest data base for protein-membrane interactions. While there are now structural data on the heptad transmembrane structures of representatives of several heptad families, here we show that fresh insights into global and some local chemical trends in GPCR properties can be obtained accurately from sequences alone, especially by separating the extracellular and cytoplasmic loops from transmembrane segments. The global mediation of long-range water-protein interactions occurs in conjunction with modulation of these interactions by roughened interfaces. Hydropathic roughening profiles are defined here solely in terms of amino acid sequences, and knowledge of protein coordinates is not required. Roughening profiles both for GPCR and some simpler protein families display accurate and transparent connections to protein functionality.

## 1. Introduction

It has long been clear that protein dynamics and functionality are determined by a combination of short- and long-range interactions. Our knowledge of short-range interactions was based initially on short amino acid sequence similarities coupled to functionality in homologous families, but recently much more precise knowledge of short-range interactions has become available primarily from crystal structure determinations of wild or modified proteins either in isolation or in combination with other proteins relevant to their functionality. Crystal structure determinations have become so successful that in recent years an annual key word search of the Science Citation Index on protein* AND structure* has produced more hits than a similar search on protein* AND sequence* (see Fig. 1).

Protein dynamics and functionality can be compared to a much simpler inorganic problem, that of the dynamics of lattice dislocations (analogous to protein chains) and their interactions with impurities (analogous to mutations in proteins) in inorganic crystals. This relatively simple network problem has been studied in great detail because of its importance for the plasticity of metals and effects on semiconductor electronics. These studies have shown [1] that short-range dislocation core interactions and long-range elastic fields contribute about equally to dislocation formation energy, while dislocation interactions are determined primarily by long-range elastic fields. The size of the dislocation core is defined by the deviation of the strain field from its long-range behavior. There is one respect in which this dislocation analogy is misleading: most dislocations that have been studied in detail (for instance, in prominent network materials like Si) are embedded in stable hosts. If the glassy network host is marginally stable elastically (as proteins with their multiple conformations are), then long-range elastic fields could dominate short-range interactions and play a decisive role in functional trends.

Interactions of proteins *in vitro (vivo)* are mediated by water, which itself is expected to determine a characteristic length scale for the protein-water interface. This length could be associated with roughening of this interface, which should be longer than an individual amino acid size because of the relative weakness of the water-protein interaction. In the presence of two length scales diffusion becomes anomalous and may lead to self-organized criticality [2]. What makes proteins special is that their hydropathic interactions have been quatntified, so that they appear to be the archetypical realization of the concept of self-organized critical (SOC) networks [3]. SOC is the obvious concept for describing any system that has been optimized, especially with respect to long-range, highly cooperative interactions, such as conformational changes. Near optimized extrema, all properties depend on power laws, whose exponents can be obtained from their slopes on log-log plots. The problem has been to find a "handle" which could be used to quantify SOC most simply in proteins, by using such exponents.

Apparently such a “handle” has been discovered [3], based on amino-acid hydrophobicity, as quantified by statistical studies of trends in solvent accessible surface areas (SASA), as determined from classical Voronoi partitioning. The unexpected feature of this seminal discovery is that these areas decrease with fragment length in protein fragments containing (2N + 1) amino acids according to power laws in (2N + 1) ($1 \leq N \leq 17$) with different (centered, amino-acid specific) exponents. These exponents, derived from bioinformatic scans of 5526 high-resolution protein fragments, show that each aa induces long-range changes in local backbone curvatures (which are smaller(larger) for hydrophilic(phobic) residues, which are respectively exposed or buried in globular proteins). Because the SOC-based MZ hydrophobicity scale is dimensionless, free of adjustable parameters, and virtually exact, it offers unprecedented transparency, reliability and transferability, the three most valuable qualities of a microscopic method.

This discovery [3] immediately suggests SOC, but at the same time it was regarded skeptically by the biophysics community, the reason being that historically there have been many different definitions of hydrophobicity (although none of the earlier ones involved evolutionary factors explicitly through exponents). In previous papers on scaffold, transport packaging, and lysozyme *c* [4,5], this skepticism was confronted with a series of successes in connecting long-range interactions with protein functionality. These successes occurred primarily in the context of hydrophobicity profiles $\langle \Psi W \rangle$. Here the SOC hydrophobicity curvatures Ψ were smoothed over rectangular windows of length W, chosen to describe the secondary structures of interest, or simply with W = 3 to describe immediate contextual effects.

Before proceeding further, there is an important general point. Short-range amino acid interactions have been studied in great detail, especially in structural studies, including mutations at key contact points or conserved sites. Does this mean that all the other amino acids (which typically constitute 90-95% of aligned “exact" amino acid sequences, or 80% if a weak “similarity” criterion is used) in a protein function only as spacers? If so, why does this content change so much from one species to the next, for example, why are only 72 sites conserved in human lysozyme *c* compared to Hen Egg White and 58 are mutated, when the $C_\alpha$ coordinates of these two lysozymes are structurally superposable to 0.65 A [5]? It turned out that the significant differences in $<\Psi W>$ between species occurred only in certain segments of lysozyme *c*, which could be identified by using $<\Psi 3>$ profiles, and these correlate well with trends in both enzymatic and antimicrobial properties [5], neither of which are explicable using structural superpositions.

Here we will see that hydropathic roughening profiles provide another handle on long-range interactions, that could be as useful for the latter as short sequence similarities and crystal structures are for exploring short-range interactions. The analysis covers briefly many proteins, as it is designed to explore the potential of roughening analysis. In each case, many more detailed analyses are possible within protein families and superfamilies. Throughout the analysis, we can identify some features which are common to similar structures, while others vary, and these latter variations usually have a simple functional interpretation.

The emphasis in this work, as in previous papers in this series, is on connecting protein aa sequences with known structures which differ within a subfamily (for the same species), or between different species, or as mutated, with changes in functionality. Traditionally there has been a large literature that has addressed the sequence-structure (folding) problem, with little or no regard to functionality. While the sequence-structure problem remains unsolved, the advent of a very large and rapidly increasing number of known structures (Fig. 1) has expanded the sequence-functionality problem, while gradually reducing the sequence-structure problem. The reader who is interested in comparing the present sequence-functionality approach to sequence-structure (folding) approaches could consider some recent examples of the latter [7,8]. These general folding approaches tend to utilize randomized aa sequences as benchmarks or controls for extracting structural information. Our method does not utilize such constructs, which are unnecessary in the presence of the nearly exact SOC-based hydrophobicity scale.

## 2. Roughening Profiles: Definitions, Rules and Benchmarks

Because proteins are confined to an exponentially complex SOC amino acid subspace designed by evolution, there is no derivation or definition of roughening profiles obtainable by polynomial methods, as is possible for continuum models that contain no chemical information [2]. Our definition proceeds as follows. From the protein amino acid sequence ($i$) ($i$ = 1,…,M) we construct ($i$,N) sliding window arrays <$\Psi_\alpha(i)$W(N)> with $\alpha$ =1,2 by averaging $\Psi_\alpha(i)$ from $i$-N to $i$ + N. Here $\alpha = 1$ corresponds to using the long-range MZ SOC hydrophobicity scale, and $\alpha = 2$ uses the KD scale based on transference energies from water to organic solvents; both of these scales are listed in [5]. These arrays encounter edge problems when $i - N < 1$. We may choose to avoid these problems by shortening the sequences so that they always begin with $i - N = 1$, and end with $i + N = M$. The resulting <$\Psi_\alpha(i)$W(N)> arrays are trapezoidal rather than rectangular, but as we seldom consider sequences with M so small that N/M is appreciable, these finite-size end limitations appear to be incidental.

As discussed before, the MZ scale [3] is based on the evolution of solvent accessible surface areas (SASA), as determined from classical Voronoi partitioning. The unexpected feature of bioinformatic scans of 5526 high-resolution protein fragments is that these areas decrease with fragment length in protein fragments containing (2N + 1) amino acids according to power laws in (2N + 1) ($1 \leq N \leq 17$) with different (centered, amino-acid specific) exponents $\Psi_1(i)$. These exponents, derived from bioinformatic scans of 5526 high-resolution protein fragments, show that each aa induces *long-range* changes in local backbone curvatures (which are smaller(larger) for hydrophilic(phobic) residues, which are respectively exposed or buried in globular proteins). The KD definition $\Psi_2(i)$ uses energies of transferring amino acids from water to organic solvents, which involve only short-range forces. Had we defined $\Psi_\alpha(i)$ merely in terms of fragment averaging over observed SASA with N = 0, we would have obtained the currently most popular hydrophobicity scale which mixes short- and long-range forces, with a correlation coefficient intermediate between that of $\Psi_1(i)$ with $\Psi_2(i)$, as discussed in [5]. The differences between different scales become important for long-range interactions with $W \geq 9$.

The hydrophobicity profiles <$\Psi_\alpha(i)$W(N)> exhibit oscillations, and these oscillations are often quite similar for different species over large parts of a given protein, but it was shown in [4-6] that occasionally there are systematic differences over long segments, and that these systematic differences correlate extremely well with protein functionality. In retrospect these strong correlations could have been expected, as the conformational changes that determine protein functionality involve long-range forces, with the changes in short-range interactions limited by secondary structures (helices and strands) whose main role is to stabilize the functional units during conformational changes. Increasing N smooths the oscillations of the hydrophobicity profiles and reduces their amplitudes, which are a measure of interfacial water-protein roughening.

The next step is to exploit a well-known property of power functions: a power function of a power function is itself a power function [$(x^{\alpha})^{\beta} = x^{\gamma}$, with $\gamma = \alpha\beta$]. This means that we can measure the amplitude smoothing by taking its variance as a function of W = 2N + 1, and obtain estimates of new exponents by studying

$$\mathrm{VAR}(W) = \sigma^2(W) = \Sigma(<\Psi_{\alpha}(i)W(N)> - <<\Psi_{\alpha}(i)W(N)>>)^2/(M-2N-1) \quad (1)$$

If these amplitude fluctuations are random, when we plot logVAR against logW, we should see a linear plot with slope $-\delta = 1$ as a first approximation, because this is what is predicted by the Central Limit Theorem. To check this idea, one can download a string of random numbers $\varphi(i)$ from a Web-based random number generator, use EXCEL to calculate the smoothed array $<\varphi(i)W>$ and VAR($<\varphi(i)W>$), and plot logVAR($<\varphi(i)W>$) against log W. One finds, as expected, $-\delta = 1$, with a finite-size correction that depends on $a/M^{1/2}$, with $a \sim 1$. The nature of the finite-size correction is known to mathematicians as the Berry–Esseen theorem. The corrections are small for sequences of 50 or more numbers, and will not be significant in our calculations. Our real interest is in identifying those features of protein amino acid sequences that depend on the differences between VAR(W) profiles that vary from one protein family or superfamily to the next, for sequences of the same or similar lengths. The alert reader will have already realized that a very attractive feature of roughening profile analysis is that it contains no adjustable parameters specific to a given protein, so that whatever properties are identified in VAR(W) are genuine results of evolution. Another very attractive feature is its high resolution, which far surpasses anything obtainable by spatial superpositions of known structures [6], or molecular dynamic studies of "flickering" water-protein interactions [9].

From the viewpoint of total energy calculations, it might appear that because water-protein interactions are weaker than carbon backbone and molar volume-dependent amino acid side group packing interactions, long-range water-protein interactions will have imperceptibly small effects on protein sequences. However, the packing interactions are short-range, and while they produce the hydropathic oscillations mentioned above, these die out for large W, leaving mainly water-protein roughening interactions. The packing interactions are so strong that the energies associated with them change little with tertiary conformations. The situation here is analogous to that described by the Huckel π electron theory of polycyclic hydrocarbons. The strong coplanar σ short-range interactions have little effect on most chemical properties, which are determined by the weaker π out-of-plane long-range polarization interactions.

Broadly speaking, hydropathic roughening profiles are dominated by hydrophobic mismatch, and this can have several effects. Where the water density is low, the packing interactions are less screened, and the secondary units such as α helices are more stable. If stability is not crucial to functionality, then hydropathic smoothing can be favored by evolution. For given protein families or superfamilies, these trends are easily recognized. By exploring them in a large number of cases, we can become familiar with the nature of the contribution of water-protein interactions to conformational interactions.

### 3. Repeat Proteins: Scaffolds and Transporters

HEAT and ARM repeat protein families have many attractive features that are especially suitable for studying long-range water-mediated interactions. They have large SASA's that are separable into patches with distinct functionalities that are easily recognized with our SOC order parameters, much like liquid crystal domains. Their secondary structures are especially simple, being composed almost completely of helical arms connected by short loops. PR65/A (588 residues, 15 repeats, PDB code 1PRO) has a very simple spiral (coiled coil) structure, each 39-residue repeat consisting of two equally long helices A and B in an L-shaped conformation connected by short, three amino-acid (aa) turns, which define a binding groove; this simple structure (77% helical) functions as a scaffold, supporting a regulatory protein (repeats 2-7) and a catalytic subunit (repeats 11-15), separated by an interfacial hinge (repeats 8 and 9). Previously [4] this structure was analyzed by calculating its hydrophobic plasticity, defined in terms of adjacent harmonic inter-repeat interactions. The water-mediated contribution to these interactions is easily recognized, because successive repeats not only have equal lengths, but they are also marked by many conserved sites. However, this geometry is very special and is most favorable for PR65/A, so this interaction was useful only in this case, where it was clearly more pronounced for the MZ scale than for the KD scale.

Here we return to PR65/A, and use it as an example of interfacial roughening, shown for the MZ scale in Fig. 2(a) and for the KD scale in Fig. 2(b). The protein has been divided into three sections, helical repeats (1-6) [bound to the regulatory unit], (8-9) [interface] and (11-15) [catalytic unit], with the boundary repeats 7 and 10 omitted to enhance the roughening separation of the three sections. The reader should see at first glance that the log-log plots are nearly linear, reflecting a considerable degree of short-range "random" disorder, with slopes of order -1. However, these features merely show that strong packing of molecular volumes dominates weak hydropathic interactions for small W. If we examine Figure 2 more closely, we discover breaks in slope at N = 4 and W = 9. If one wishes, one could associate this break with doubling of the α helix 3.6 residue periodicity, except that the same break is found regardless of secondary structural content. It might be better to say that there is a universal elastic length scale associated with water-protein interfaces, and that when ~ 70% α helices are present to stabilize the structure, their pitch is adjusted to fit this length scale.

The analysis of [4] led to the conclusion that the MZ scale explains the functionality of the PR65/A scaffold better than the KD scale, because the hydroplasticity peaks at repeats 4, 8-9 and 13 in Fig. 1 of [4] were sharper with the MZ scale. The central interfacial repeats 8-9 show rapid hydrosmoothing (sub-linear reductions in VAR(W)) with increasing W for the MZ scale (Fig. 2(a)), relative to the binding repeats 1-6 and 11-15. This is easily understood: the smoothing enables the interfacial repeats to move with less hydrodrag. However, the opposite is true with the KD scale, as shown in Fig. 2(b). This means that the KD scale, because it does not include long-range interactions, actually generates spurious above-average drag on the interfacial region (the KD scale is intrinsically noisy, and the background noise does not decrease so rapidly as it should to explain the greater hydroflexibility of the interfacial region relative to the binding regions). This analysis is continued elsewhere [10] for other transport and packaging proteins, as well as lysozyme *c*. We now turn to transmembrane GPCR.

## 4. GPCR Proteins: Global Roughening Profiles of Family A, Rhodopsin, Adenosine, Adrenergic and Dopaminergic Receptors

Guanine protein coupled receptors (GPCR) (membrane signal transducers) were first identified from sequence similarities or homologies as possibly the largest protein superfamily, with 800+ human and 3000+ total GPCR sequences listed so far [11]; GPCR's make up roughly 3% of genes in the human genome. Hydropathicity analysis revealed seven hydrophobic stretches corresponding to transmembrane spanning alpha-helices (TMI-TMVII), connected by interior cytoplasmic and extra-cellular loops [11]. The near-universality of TM heptad ring GPCR structures could be the result of the stability of an odd number of helices against bimodal fragmentation, with five helices usually yielding too small a ligand binding cavity, and nine being unstable.

The first high resolution structure of a G-protein-coupled receptor (GPCR) was that of bovine rhodopsin (optical signaling), but most important GPCR signals are chemical. On any given day you turn over in the adenylate cyclase cycle your body weight equivalent in adenosine triphosphate (ATP), the principal energy currency of the cell, so the recent discovery of the structure of adenosine transmembrane receptors has stimulated great interest [12,13]. Because adrenalin is the body's strongest native stimulant, the discovery of the structure of adrenergic GPCR is equally important pharmaceutically [14]. For the reader's convenience Fig. 3 shows ligands for four classes of receptors that are discussed here. Except for the similarity between adrenalin and dopamine, there are large differences in morphology between the ligands, which may be related to topological differences in hydrophobicity profiles of receptors. Although the locations of these ligands when bound to their respective GPCR's are known in *ex vitro* crystal structures, the latter exhibit mainly short-range interactions which may not be fully representative of typical interactions *in vivo*, especially of transition states. Because GPCR functionality depends on long-range, water-mediated interactions as well, we can look for patterns in the latter which exhibit interesting correlations with receptor functionality not immediately obvious from structural or mutational studies.

All observed GPCR structures have confirmed the heptad helical structures anticipated from sequence analysis, so the next question is what are the major structural differences, and how do these, together with sequence differences, relate to functional differences? [13] attempted to answer this question, and confirmed that the major sequence similarities occur in transmembrane (TM) subsequences, at a level of about 18%, as aligned relative to prototypical transmembrane proteins using structural similarities. This is large enough to be easily recognized, but is it useful? In particular, could aligning known helical subsequences be useful in assigning GPCR's to be used as overall templates for unknown protein structures? Fourteen of the latter were studied [13], and their TM subsequences optimized with respect to each TM subsequence of a known structure, with inconclusive results. Generally the maximum TM subsequence similarities of the structurally unknown proteins exhibited small variations between the five known putative TM subsequence templates. Within the context of these small variations, functionally plausible results were obtained; for example, overall dopamine D2 receptors sequences matched adrenergic receptors most closely (adrenalin, Fig. 3(a), and dopamine, Fig. 3(d), are similar), rather than rhodopsin or adenosine receptors. However, when specific receptor structural features were included in the analysis, the final template suggestions for dopamine D2 receptors ([13]'s Table 7) inconclusively contained five adrenergic and three rhodopsin preferences.

The difficulty of comparing sequence and structure similarities in this reductionist way, through studying spatially their most stable elements separately, even with five known structures, was implied by the observations made in the original structure reports, to the effect that some of the largest differences between the best studied GPCR's occur in the extracellular loops [15,16]. These interact strongly and completely occlude the rhodopsin binding site from solvent access. In comparison with rhodopsin the extracellular region of the adrenergic receptors is very open, as is that of adenosine receptors, so open that it is structurally stabilized by four disulfide bridges. At the same time, the ligand binding pockets of rhodopsin and the adrenergic receptors are similar, but that of the adenosine receptor is quite different, having rotated to be normal to the plasma membrane plane, and much shallower, with almost half of the ligand completely exposed to bulk solvent. Finally, a short-range feature of the intracellular loops in rhodopsin, which had been assumed to be common to all GPCR's and was called the ionic lock (between two conserved proximate residues on TMIII and TMVI), is remote in the adrenergic and adenosine GPCR's.

## 5. Rhodopsin

GPCR-ligand interactions are best understood for rhodopsin. Rhodopsin has long been the paradigmatic GPCR, because optoelectrochemical transduction can be studied in many ways. The retinal-containing rhodopsin pocket [17] and opsin R* - transducin $G_t$ complexes [21] are known from structural studies. These showed that the cytoplasmic side of the TM5/TM6 helix pair (corresponding to the cytoplasmic loop C3 of the 7TM bundle, here also called 5.5) forms a mitt-like structure in which $G_t$ is held. The lengths of loops C1 (~10) and C2 (~20) are nearly constant, while the lengths of C3 vary widely among GPCR. Later we will correlate these C3 length variations with appropriate hydrophobic factors derived from the MZ SOC scale. Note that the length of C3 is smallest for vertebrate rhodopsins (22 amino acids,increasing to 34 for squid), so vertebrate rhodopsins can be regarded as basic not merely because their crystal structures were discovered first, or because their ligand-binding state is best understood, but also because their cytoplasmic interactions are simplest.

It has been argued that rhodopsin kinetics are primarily short-range and can be understood in terms of a reaction channel which is ca. 70 angstrom long and between 11.6 and 3.2 angstrom wide [18]. The narrow constrictions within the channel must stretch to allow passage of the retinal beta-ionone-ring (see Fig. 3(a)). Here we will use rhodopsin as a base for analyzing chemical hydrophobicity trends which are likely to be important for receptor-ligand interactions in the cytoplasm, even when these interactions are remote and are mediated by hydroelastic waves without long-range ligand transport. The location of the rigid antagonist carazolol in the $\beta_2$-adrenergic receptor is very similar to that of retinal in rhodopsin [19], but the pocket in adenosine receptors is different, partly because the adenosine connectivity is altered by four disulfide bridges in the extracellular domain [20], which is stabilized by disulfide bonds [21].

Hydrophobic analysis of the extracellular and cytoplasmic C3 loops sheds further insights into these long-range interactions (next section).

Broadly speaking, water penetrates deeply into membrane proteins [22], which constitute 1/4 of all known proteins [23], and the concept of water as a mediator of conformational dynamics in signaling is gaining ground [24]. Transmembrane interactions occur on a large scale, and lipid-protein interaction in membranes is governed by hydrophobic mismatch [25]. Physiologically relevant variations in both the (hydrophobic helix length)/membrane thickness ratio and cholesterol levels influence transmembrane helical orientation [26] and receptor aggregation [27]. Hydrophobic roughness is a tool that is ideally suited to studying large-scale, protein-specific trends in collective helical interactions separately from medium-scale interactions involving individual TM transmembrane helices. An extreme example is presented by the large differences between mammalian rhodopsins and bacteriorhodopsin, which present serious structural and sequence-matching problems, although both are helical heptad structures [28].

### 6. Separated Roughness

In repeat proteins and lysozyme *c*, contextual hydrophobicities involving short windows with N = 1 and W = 2N + 1 = 3 yielded useful information through their hydroprofiles [4-6]. However, because the MZ scale is so accurate, one can also use values of W that are even larger than helical lengths. In the past it was suggested that the secondary TM structure itself is best seen with N ~ 9 and W ~ 19, matching longer helical lengths [28]. Here we explore hydroroughness over an even longer range, $1 \leq N \leq 23$ and $1 \leq W \leq 47$. Still larger values of N and W appear to yield noisy results, possibly because the MZ scale itself explored N values only up to 17.

The special geometry of transmembrane proteins can be exploited by modifying the general definition of roughness given in Eqn. (1), based on hydropathic quadratic deviations of $<\Psi_{\alpha}(i)W(N)>$ from its average value $<<\Psi_{\alpha}(i)W(N)\gg$ for the whole protein (its variance). Instead one calculates $<<\Psi_{\alpha}(i)W(N)\gg$ for the three transmembrane, extra-cellular and cytoplasmic regions of each protein separately, and then uses these three separate averages to calculate hydropathic quadratic deviations of $<\Psi_\alpha(i)W(N)>$ for the entire protein. In effect, one assumes that each of the three regions defines a local environment, and roughness is determined by variance fluctuations relative to that local environmental average. Thus the transmembrane segments tilt relative to each other, and their motion relative to the connecting loops is included only implicitly through the length of the averaging window W. The latter can be optimized, as we will see.

Separation into three regions has many advantages over the simpler variance method where the fluctuations are calculated relative to the protein as a whole, as then variations in the lengths of connecting loops (which are especially large for the cytoplasmic terminal segment) have much less effect on the roughness of the transmembrane region. However, special care is required in handling the cytoplasmic terminal segment. Generally one would guess that the effects of very long terminal segments should be restricted by cutting off that segment after it has reached about twice the length of the cytoplasmic 5.5 segment. In rhodopsin proteins this works well (terminal, 39 residues, 5.5 loop, 22 residues). This is also true of many other opsins, but the Uniprot P29274 sequence for Adenosine A2a receptors contains 122 cytoplasmic terminal residues, compared to 36 for the cytoplasmic 5.5 segment. The reported crystal structure (PDB 3EML) is cut off at only 28 cytoplasmic terminal segment residues. In practice all terminal cutoffs between 1-2 lengths of the 5.5 cytoplasmic loop yield similar results for global protein functionality, and we have chosen to cut off GPCR terminal segments at the rhodopsin ratio 39/22, for example, in Adenosine A2a at 64 ~ 39x36/22 residues. This is the only GPCR studied here where a cutoff is needed.

## 7. Rhodopsin Details

It turns out that among GPCR proteins rhodopsin is hydropathically special at large window lengths W, for several reasons, but before discussing these it is interesting to explore the difference between vertebrate (348 amino acids) rhodopsin GPCR and the primitive ion pumps (262 amino acids) bacteriorhodopsin (proton pump) [30] and (274 amino acids) halorhodopsin (Cl pump). As shown in Fig. 4, the most interesting hydropathic roughening profile differences between these rhodopsins occur near W = 47, which represents the effect of water on helical-loop-helical interactions – in other words, global coupling of rhodopsin to cytoplasmic proteins. Halorhodopsin is a promising high-performance genetically targetable optical neural silencer [31]. The largest evolutionary structural differences between the primitive ion pumps and vertebrate rhodopsin occur in the lengths of the loops (average loop length in vertebrate rhodopsin is 22 residues, compared to only 10 (14) in halo(bacterio)rhodopsin, while the average TM length remains at 24 residues). Note that the smoothest long-range (W = 47) rhodopsin is bacteriorhodopsin, even though halorhodopsin has shorter loops. The increased roughness of halorhodopsin reflects the larger size of the Cl ion relative to the proton.

Given that human rhodopsin is much more evolved than the primitive ion pumps, why are the latter hydropathical profiles smoother in Fig. 4? It appears that the primitive ion pumps are not constrained by their coupling to guanine proteins, in other words, the demands placed on GCPR by their signaling functions requires a more complex (and rougher) hydropathic structure. We will see later that many adenosine, adrenergic and dopaminergic receptors are smoother than rhodopsin in the midrange 3 < W < 21-25, which reflects the complex nature of the retinal-rhodopsin interaction. At longer range, W > 25, rhodopsin smooths more rapidly than other opsins, reflecting the crucial role played by water in correlating optical signaling of its heptad transmembrane helices.

[10] lists salient roughening features for many rhodopsins, and it contains many instructive trends measured by long-range hydropathic interactions. These trends are significant because the MZ SOC hydrophobicity scale is apparently much more accurate in quantifying relative protein properties than any other scale, and certainly much more accurate than total energies obtained from the classical force fields used in MD simulations. [10] contains roughness profiles for many species calculated both with the separation definition and the simpler overall variance method. While the separation method nearly always gives smaller values, the chemical trends of the two methods are quite similar for rhodopsin, where the average loop length is less than the average transmembrane length.

Here we focus on only two species, human and rat, because their apparently small differences are easily identified with the SOC MZ scale, and these differences are pharmaceutically important. As shown in Fig. 5, with the unified roughness definition, the human-rat profile differences peak at W = 15 at about 8%, while with the separated definition they increase up to W = 15, where they flatten out and remain nearly constant at about 5%.

8. **Adenosine Receptors**

There are four adenosine receptors in humans. Each is encoded by a separate gene and has different physiological functions (as given by Uniprot), with $A_1$ and $A_{2A}$ playing central roles in the heart, regulating oxygen consumption and blood flow, while the $A_{2A}$ receptor also has broader anti-inflammatory effects throughout the body. The $A_{2B}$ and $A_3$ receptors are located mainly peripherally and are involved in processes such as inflammation and immune responses, while central $A_1$ inhibits (peripheral $A_{2B}$ stimulates) adenylate cyclase activity (cyclic adenosine phosphorylation). Adenosine receptors are major targets of caffeine. After more than three decades of medicinal chemistry research, a considerable number of selective agonists and antagonists of adenosine receptors have been discovered, and some have been clinically evaluated, but only Regadenoson, which is an $A_{2A}$ agonist that is a coronary vasodilator, has received regulatory approval [31]. Regadenoson has a 2-3 minute biological half-life, as compared with adenosine's 30 second half life, and it has an additional stabilizing $C_3N_2$ ring attached to the $C_4N_2$ ring of adenosine (Fig. 3(c)).

The sequences of the four adenosine receptors in humans all have similar lengths from the N terminal to the end of TM7. The hydropathic roughening profiles of the four adenosine receptors exhibit systematic trends, as shown in Fig. 6. Roughening separates central $A_1$ and $A_{2A}$ as a more flexible subgroup at higher pressures for short-range interactions (W < 9), compared to more rigid peripheral $A_{2B}$ and $A_3$ subgroup receptors, as one would have expected. For W > 11 long-range interactions, the four receptors regroup into two different subgroups, "smooth" $A_1$ and $A_3$, and "rough" $A_{2A}$ and $A_{2B}$ receptors. This long-range grouping is consistent with stimulation (inhibition) of adenylate cyclase by rough $A_{2A}$ and $A_{2B}$ (smooth $A_1$ and $A_3$) receptors, and the 56% sequence similarity of $A_{2A}$ and $A_{2B}$ receptors [32]. The "rough" $A_{2A}$ and $A_{2B}$ receptors can interact effectively with the ionic phosphorylation cycle, while the "smooth" $A_1$ and $A_3$ receptors merely obstruct it.

Is this striking short-long range regrouping merely an artifact of our analysis? This question can be answered by comparing the human roughening profiles of Fig. 6 with the similar rat roughening profiles shown in Fig. 7. Comparison shows that the profiles for A1, A2b and A3 are little changed, but the rat profile for A2a is roughened relative to human A2a so much that the crossover noted in Fig. 6 has disappeared. The additional rat profile roughening occurs only in the separated definition, and is maximized at 30% for W = 13, while in the unified definition rat A2a is reversed as a few % smoother than human A2a. The rat/human Blast sequence identities are 82% for A2a and (94, 86, 74) for A1, A2b and A3. It appears that the roughening profiles are a much more informative measure of evolutionary specialization than sequence identities.

We expect that human smoothing is a refinement made possible by evolution. The source of the increased rat roughening is readily identified specifically in the W13 profile as a region in the second extracellular loop EC2 near TM IV centered on rat 146K147D, three residues beyond the rat disulfide bond involving Cys 143 (142N143C144S). The human smoothing occurs near the conserved Cys146 (145N146C147G) that bridges the loop with the TMIII; this conclusion was also reached using a graphical method with multiple sequence alignment of the TM domains [33]. Clustalw2 yields an 81 score for similarity of human A2a with rat A2a, compared to (94, 86, 73) for A1, A2b and A3, yet the overall (long-range!) human roughening A2a profile is drastically different from that of rat A2a.

The above long-range difference, based on all residues and inaccessible to multiple sequence alignment of like residues, has important consequences; for example, EC2 of human adenosine A2a was used to immunize mice for production of Adonis, an IgM monoclonal antibody [34], which binds to the loop near TMV. There EC2 is much more flexible in humans than in rats or mice because of hydrosmoothing in the part of EC2 between the disulfide bonds and the antibody binding site. Many studies of interactions between adenosine and dopamine receptor antagonists with different selectivity profiles have relied on measurements on rats [35], but comparison of Figs. 6 and 7 suggests that the effects of A2a antagonists on dopamine receptors could be both different and larger for humans. Human studies show caffine correlations between adenosine A2a and dopaminergic D2 receptors [36]. Abnormal expression of adenosine A2a and dopaminergic D5 receptors may have a significant role in the pathogenesis of obsessive-compulsive behavior [37]. Immunosuppressive signaling via the A2A receptor could explain the inefficiency of antitumor T cells in the tumor microenvironment, which suggests a promising approach to enhance anti-tumor or anti-pathogen immune response [38].

**9. Adrenergic receptors**

R. Ahlquist suggested in 1948 that adrenalin can affect muscles through adrenergic receptors either by contraction (α receptors) or relaxation (β receptors). By 1964 J. W. Black had developed propranol, which blocked β receptors and decreased oxygen demand of the heart. Today adrenergic receptors are separated into three groups based on $G_q$, a phospholipidase ($\alpha_1$); $G_i$, which inactivates adenylate cyclase ($\alpha_2$), and $G_s$, which stimulates cAMP ($\beta_1$ –$\beta_3$). Recent studies [39] of docking at $\beta_2$ receptors surveyed 21 ligands (full, partial and inverse agonists and antagonists). The binding results are sensitive to the derived $\beta_2$ receptor geometry, which was close to that observed by diffraction [20].

The present analysis enables the study of the adrenergic families of α and β hydrophobic patterns without detailed knowledge of their structure, known at present only for modified $\beta_1$ and $\beta_2$ receptors. The shapes of the roughening profiles of human adrenergic receptors are qualitatively similar (Fig. 8). Using only the variance definition of roughness, human α and β groups are hierarchically well-separated in Fig. 8(a), except for the near coincidence of $\alpha_1$A and $\beta_1$. When the roughness profiles are calculated using separated averages for the three transmembrane, extra-cellular and cytoplasmic regions of each protein, one obtains a fully satisfactory separation of human α and β wild type hydrophobic profiles, Fig. 8(b). The relative variations in Fig. 8(a) are small, while in Fig. 8(b) adrenergic receptor α(2A) exhibits the greatest variation, with the largest roughness for large W, and much less below W ~ 15. α(2A) is the predominant

α(2) subtype in the central nervous system and exerts a sympathoinhibitory (hypotensive) action; on the contrary, activation of the central α(2B)-AR elicits a sympathoexcitatory response (such as seen in salt-induced hypertension) [40]. Thus Fig. 8(b) [separated averages for the three transmembrane, extra-cellular and cytoplasmic regions of each protein] is much more informative than Fig. 8(a) [simple variance].

This global primary variance boundary between human α and β adrenergic groups is addressed more locally in the next section by studying secondary V-VI cytoplasmic GPCR loops. Overall the excellent separation of human α and β groups by global roughening profiles for all length scales (Fig. 8(b)) should be considered to be a great success for theory, as it shows that muscle relaxation or contraction is determined primarily by water-protein interfacial interactions over a wide range of length scales, which are calibrated by hydropathic SOC separated variances in a length-invariant way. The overall good agreement between the trends identified by hydropathic SOC variances is consistent with a simple model for G-protein signaling based on stages of phosphorylation [41].

Because this $\alpha_1$A and $\beta_1$ adrenergic profile separation is so significant, one can examine its species dependence. By conventional similarity methods (which divide amino acids into four or five groups), there is little or no difference between human and rat α and β adrenergic families. What happens to the successful separation of human α and β groups by global roughening SOC profiles when rat sequences are used instead? Surprisingly, the $\alpha_1$A - $\beta_1$ human separation practically disappears for rat $\alpha_1$A - $\beta_1$ adrenergic families. Thus for W = 25, the [unified,separated] normalized $\alpha_1$A - $\beta_1$ differences (which must be positive for complete α and β separation) are [0.02,0.15] (human), and [-0.15,0.06] (rat)! Here we see two effects: the separated profile is better than the unified one, but at the same time there is a large evolutionary refinement of the α and β functional separation between rats and humans. Because the separated profiles are greatly superior, only these are shown in Fig. 9. One can see that the short-range separation (below W ~ 9) is similar for rat and human, but the mid-and long-range separations are effective only for evolutionarily more advanced humans. Moreover, for humans the separation is largest just for the midrange transmembrane length scale (W between 11 and 29). It is remarkable that the SOC scale is able (almost effortlessly) to resolve such important structural and functional differences over such a narrow evolutionary range.

Given the overall similarities between adrenergic human and rat separated roughening profiles for $\alpha_1$A and $\beta_1$ up to W ~ 5-7, what causes the steady drop in human $\beta_1$ separated roughening up to W ~ 15? It turns out that the cytoplasmic 5.5 rat loop extends from 249Arg to 308Glu, while the 5.5 human loop extends 11 amino acids further, from 249Arg to 319Glu. The average 249-308 rat roughening is 180, while the average 249-319 human roughening is only 70, a factor of 2.5 smaller. Between 280 and 300 the human sequence contains 9 Pro and 8 Ala, while the rat sequence contains only 4 Pro and 2 Ala. Thus the hydrosmoothing of human $\beta_1$ roughening for mid- and long-range W is due almost entirely to the addition of not only 5 Pro but also 6 Ala to the cytoplasmic 5.5 rat loop between 280 and 300, a remarkable example of large-scale evolutionary cooperative refinement. While Pro is often described as a kink inducer in transmembrane GPCR helices [41,42], here it plays a quite different role in evolutionary refinement of the crucial cytoplasmic 5.5 loop. A better explanation for hydrosmoothing by the additional 5 Pro and 6 Ala is that their average hydrophobicity is 0.141, which is very close to 5.5 cytoplasmic average over all five adrenergic receptors of $0.137 \pm 0.009$. At the same time Ala, the smallest and most flexible amino acid, compensates the rigidity of Pro, whose ring forces the CO-NH amide sequence into a fixed conformation.

The rigid Pro-rich human region 269-292 (14 Pro/24 amino acids) alone was previously shown to distinguish between human $\beta_1$ and human $\beta_2$ receptor coupling and sequestration [43], but the Pro content of 269-279 is unchanged between rat and human. The average 269-292 rat roughening is 220, while the average 269-292 human roughening is only 54 (factor of 4), a 60% evolutionary improvement over the entire cytoplasmic loop ranges of 249-308 rat and 249-319 human. However, the roughening values for the 280-300 ranges are rat, 292 and human only 25 (factor of 12 smaller, another evolutionary improvement, this time by 300%!). It would be most interesting to repeat the well-cited pharmaceutical 269-292 experiments [43] for the 280-300 range, either for human $\beta_1$ receptor coupling alone, or for both rats and humans.

## 10. **Dopaminergic receptors**

There are five dopaminergic receptors, separated into two groups, based on their enhancement (D1{53} and D5{50}) or inhibition (D2{162}, D3{109}, D4.2{101},D4.7{181}) of adenylate cyclase. The roughening profiles of these dopaminergic receptors are shown in Fig. 10. They separate nicely into enhancement and inhibitive groups, with inhibitive D2 and D3 significantly rougher than weakly inhibitive D4, which can also be said to lie in a border region between enhancement and inhibitive groups. The D4.R receptors occur with variable numbers (2-7) of 16 amino acid tandem repeats R, corresponding to $\lambda$ between 101 and 181. The D4.R repeats are nearly hydroneutral, and they have little effect on the roughening profiles.

When the D4.R repeats were first discovered, it was suggested that enhanced D4 concentrations could play a role in schizophrenia, as the combined density of D2 and D3 receptors is increased by only 10% in schizophrenia brain, while the density of dopamine D4 receptors is sixfold elevated [41]. Later authors have suggested more specifically that enhanced D4.7, compared to D4.2, could be another factor in schizophrenia, which is consistent with D4.7 being closer to D1,D5 than D4.2. Many statistical studies have not produced positive correlations between D4.7 concentrations and schizophrenia, but there is some positive evidence for correlations with attention deficiency hyperactivity disorder [42]. Bearing in mind the high level of sequence similarity in the dopamine subfamily [42], and the difficulty of measuring electrochemical neural activity, it is still possible that D4.R repeats are involved in schizophrenia.

Numerous studies have shown [46,47] that the polymorphic repeat sequence has little influence on D4 binding profiles and in the ability of the D4 receptor to block cAMP production. Roughening profiles confirm this conclusion because they provide an accurate measure of the difference between human D4.7 and D4.2, compared to rat D4 (which contains no repeats). As shown in Fig. 11, the R repeats have little effect on the hydroprofiles. Similarly small differences are seen in aligned hydroprofiles in Fig. 12.

## **11. GPCR Proteins: Family A, Hydroprints**

Could we have learned more about the differences between proteins belonging to the same family by studying the internal evolution of VAR(W) by examining squared differences of <ΨW> (relative to < Ψ>) and coarse-graining these with a second smaller window width? When this is done for human/bovine (the differences are small) and squid rhodopsin receptors, < Ψ17> and <Ψ21>, coarse-grained by $W_2 = 13$, one obtains the results shown in Fig. 15(a) and (b). The patterns shown here can be called hydroprints, and altogether they are both expected and surprising. The human rhodopsin hydroprint is surprisingly simple, as the squared deviations are concentrated in TM V, the short cytoplasmic loop 5.5, and TM VI. This is in excellent agreement with structural studies of bovine Ops*-$G_\alpha$(carboxy terminus) complexes, where all the binding occurs to TM V and VI [48], so it seems that hydroprints predict this central feature of receptor–G protein coupling geometry quite well.

On the basis of sequence similarities and conservation, it appears that the only large difference between vertebrate human/bovine and invertebrate squid rhodopsin receptors should be in the V-VI region, and this is what is seen by superposing the structures [49]. However, hydroprints tell more complex stories (Fig. 12). The squid hydroprint (b) shows strong roughening at TM I,III,IV,V,(not VI), and cytoplasmic loops 3.5 and 5.5 (C2 and C3). The invertebrate squid opsin roughening appears to be much more widely distributed. If one compares long range W = 45 profiles, one finds (in agreement with evolutionary biologists [50]) that lamprey rhodopsin is the lower vertebrate rhodopsin closest to squid rhodopsin. However, in the hydroprint TM W = 13 midrange, lamprey rhodopsin is still quite similar to other vertebrate rhodopsins, Fig. 15(c). One can say that “abrupt” invertebrate – vertebrate rhodopsin evolution occurs only on the W =17,21 membrane resonant thickness scale, while rhodopsin evolution appears to be much more nearly continuous on the shorter W ≤ 13 and longer W ≥ 25 scales. These are not new results for the vertebrate rhodopsins, but they are new for the invertebrate rhodopsins, which are not easily distinguished by structural comparisons alone [49].

Let us turn now to the adenosine family and the ionic phosphorylation cycle, by looking at the four hydroprints shown in Fig. 13. These show that there are two groups, [$A_1$ and $A_3$], and [$A_{2A}$ and $A_{2B}$]. [$A_1$ and $A_3$] resembles rhodopsin, in the sense that the two strongest roughening peaks are TM V and Cyto 5.5, while the two strongest [$A_{2A}$ and $A_{2B}$] peaks are Cyto 5.5 and Extracel. 4.5 – both the latter interactions being outside the membrane. As in Sec. 5, this grouping is consistent with stimulation (inhibition) of adenylate cyclase by [$A_{2A}$ and $A_{2B}$] ([$A_1$ and $A_3$]) receptors, and the 56% sequence similarity of $A_{2A}$ and $A_{2B}$ receptors [33].

An important feature of $A_{2A}$ is the three stabilizing disulfide bonds linking Extracel. 2.5 to Extracel. 4.5, compared to one disulfide bond for $A_{2B}$. In comparing the $A_{2A}$ and $A_{2B}$ hydroprints, it is also important to notice the difference in ordinate scales. The main change between $A_{2A}$ and $A_{2B}$ is not the enhancement of Extracel. 4.5 for $A_{2A}$, relative to $A_{2B}$ (they are actually nearly the same), but the reduction in the Cyto 5.5 peak in $A_{2B}$ (from a value similar to $A_1$ and $A_3$) by a factor of 2 (this could explain why two additional disulfide bonds are needed to stabilize $A_{2A}$). The constructed $A_{2A}$ used in the crystallographic study of an antagonist-$A_{2A}$ complex [21] had most of its Cyto 5.5 loop replaced by stabilizing lysozyme from T4 bacteriophage. By comparing the ligand binding pockets of rhodopsin and adrenergic receptor with their $A_{2A}$ construct, [22] concludes that the deep binding of the former to TM III, V, VI and VII has been replaced by shallow binding to TM VI and VII, as well as Extracel. 4.5 and 6.5. The hydroprints of $A_{2A}$ and $A_{2B}$ are consistent with this picture (TM VI and VII peaks are larger for $A_{2A}$ and $A_{2B}$ than for $A_1$ and $A_3$). It appears that the main difference between $A_{2A}$ and $A_{2B}$ is their phosphorylated interaction with G proteins through Cyto 5.5.

**Discussion**

Hydroelastic effects can dominate and explain functional differences between proteins with common folds, which differ between species and/or because of polymorphisms. These small differences are accessible to theory without the use of adjustable parameters because the MZ hydrophobicity scale is based on self-organized criticality (SOC), a universal principle that governs the structures of all proteins *in vivo* and implicitly includes all evolutionary effects, as shown by an overall map of the GPCRs in a single mammalian genome [51]. SOC implies that the most physiologically important protein-protein interactions are long-range. Here several examples have shown that these long-range conformational interactions (for instance, in the relative movement of helices V and VI upon photoactivation of different rhodopsins [52]) are stress-driven, and can be analyzed in terms of self-consistent multi-length scale responses [53] as described for GPCR's by hydrophobic variances.

Transmembrane proteins are dominated by a common structure - seven helices crossing the membrane, connected by loops inside and outside the cell [54]. There is a natural length scale - the membrane thickness - which is common to all of them, and the most important intracellular interactions involve cytoplasmic lateral length scales, which are large compared to the shorter length scale transverse to the thin membrane. In this context one can refine the usual definition of variance to include separated variances for extracellular, transmembrane and intracellular interactions. Long-range intracellular interactions, for instance, can be analyzed by comparing mutated and unmutated human and mouse proteins. This analysis is extremely economical and can be completed in hours, whereas the experiments take years! This is important, because protein engineers must make inter-protein comparisons to decide what to synthesize next, after an initial high throughput statistical screening has identified some promising possibilities [55]. Evolution is built into many comparisons of protein sequences, and it can be used to evaluate alternative directions for pharmaceutical research.

Although the analysis of various members of the GPCR superfamily given here has been detailed, the central conclusion is simple. Because of self-organized criticality, the MZ hydrophobicity scale can be used to construct a variety of profiles with robust hierarchical content from amino acid sequences alone. Again, because of self-organized criticality, compared to all other approximate biophysical scales, the MZ hydrophobicity scale appears to be nearly exact for all practical purposes. This makes it especially useful for not only for revealing unexpectedly large differences (Atlantic and Pacific salmon), but also and especially for studying small differences (for instance, between mouse and human) that have often proved to be inaccessible to other approaches, even to powerful and detailed multiple structural studies.

The central limitation of analysis based on the MZ hydrophobicity scale is that its effectiveness relies on protein sequence evolution. This means that the short- range details of ligand-protein interactions are inaccessible to MZ hydrophobicity profiling, because ligands are not part of the protein "universe". However, there are important pharmacological problems associated, for example, with humanizing murinal antibodies outside the immediate range of epitope interactions, which lie entirely within the protein "universe". We plan to discuss these antibody problems elsewhere, as it is already known that antibody activity can involve both short-range structure-specific protein epitope recognition and long-range nonspecific hydrophobic interaction with the viral lipid membrane [56].

**Methods**

The hydrophobicity scales are listed in [6]. The calculations utilized an EXCEL macro built by Niels Voorhoeve.

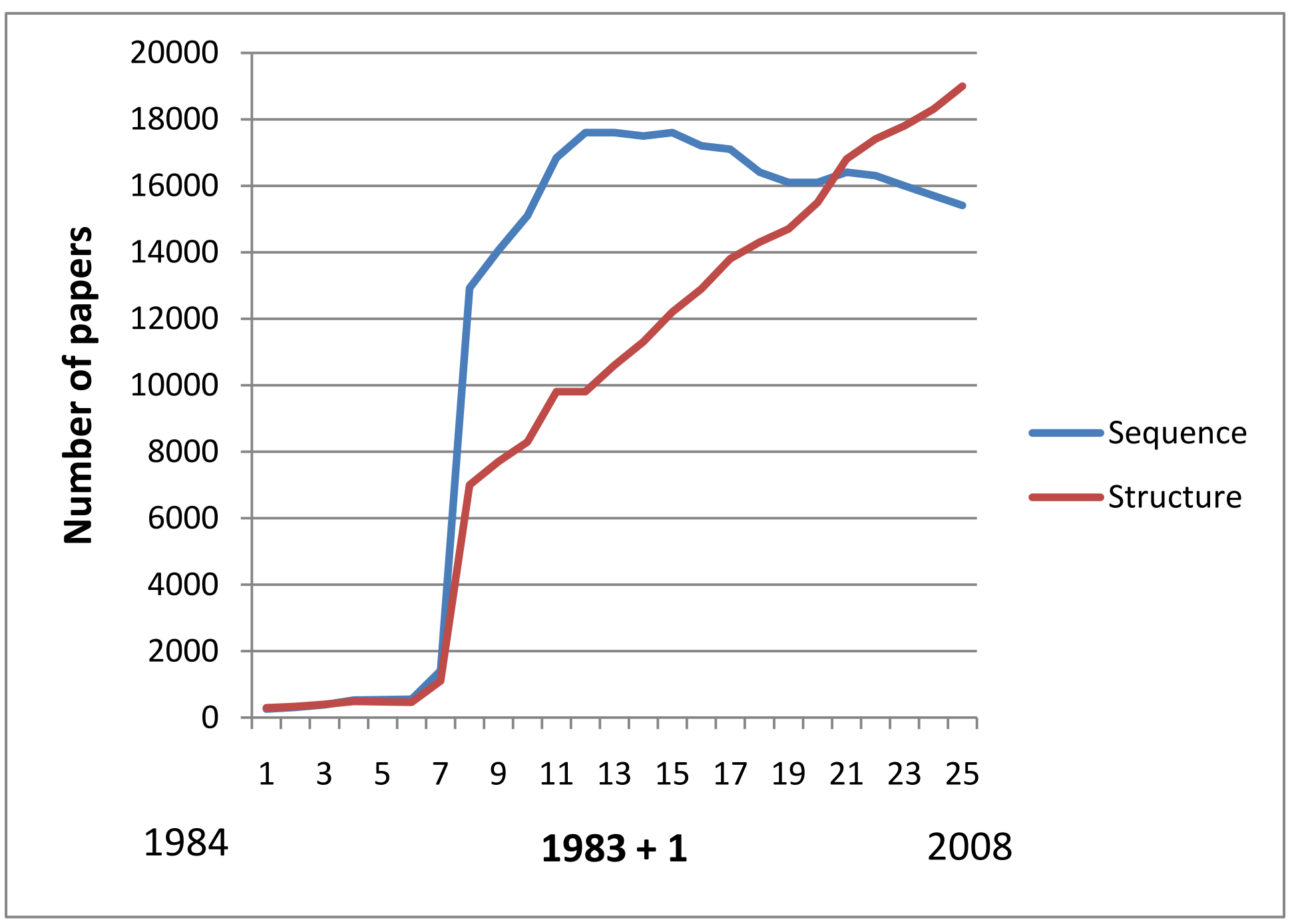

Fig. 1. Dynamics of papers with protein* AND sequence* compared to protein* AND structure*. The steady growth of protein structure studies may itself reflect the larger number of crystalline samples available, in part due to computer-automated crystal growth.

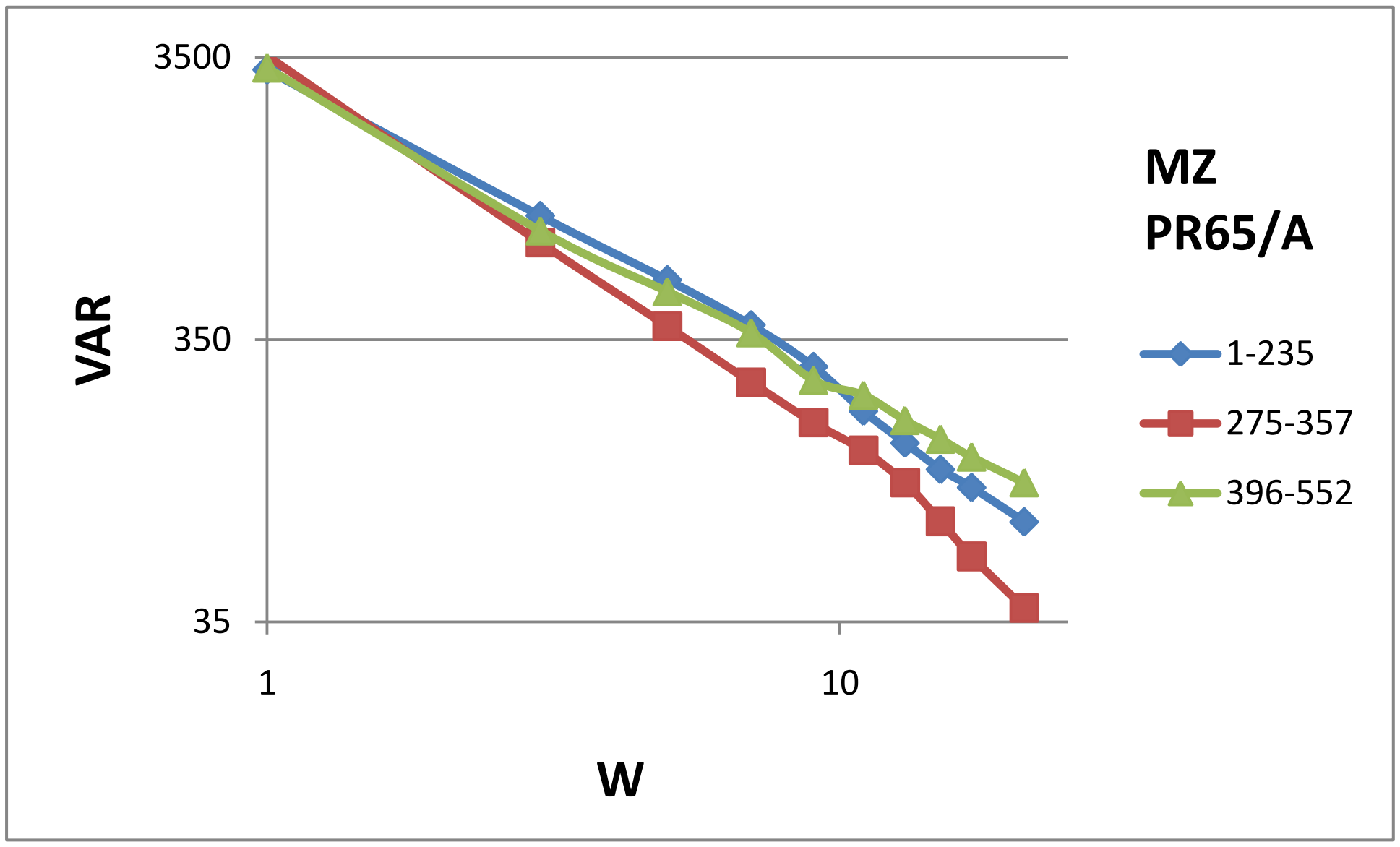

Fig. 2(a)

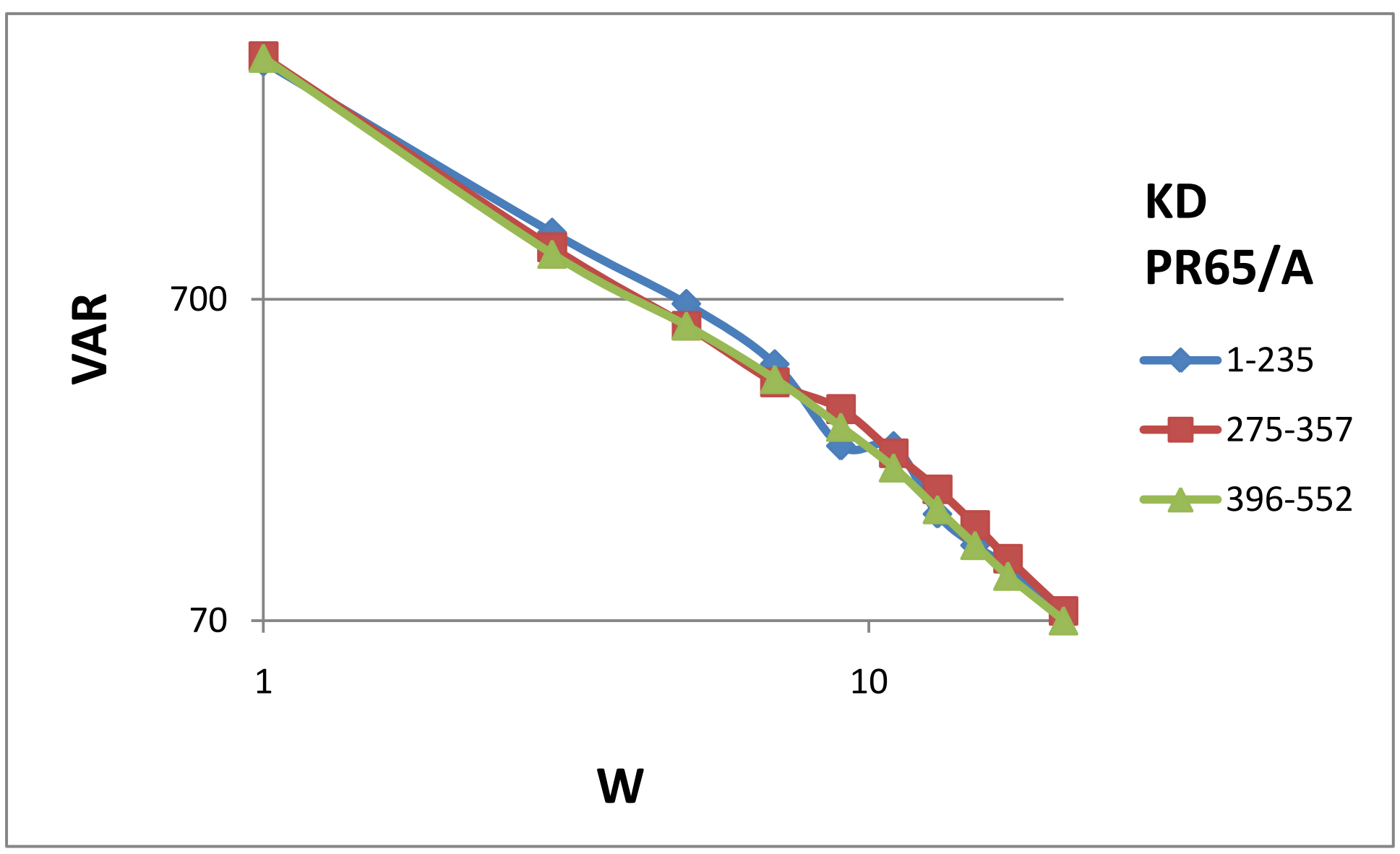

Fig. 2(b)

Fig. 2. The roughening profiles of the three sections of the scaffold repeat PR65/A.Thes three sections are clearly separated for the MZ SOC scale in Fig. 2(a), but the separations are submnerged In noise for the KD scale in Fig. 2(b). Because the lengths of the A and B arms are ~ 20 residues, the window lengths are cut off at W = 21.

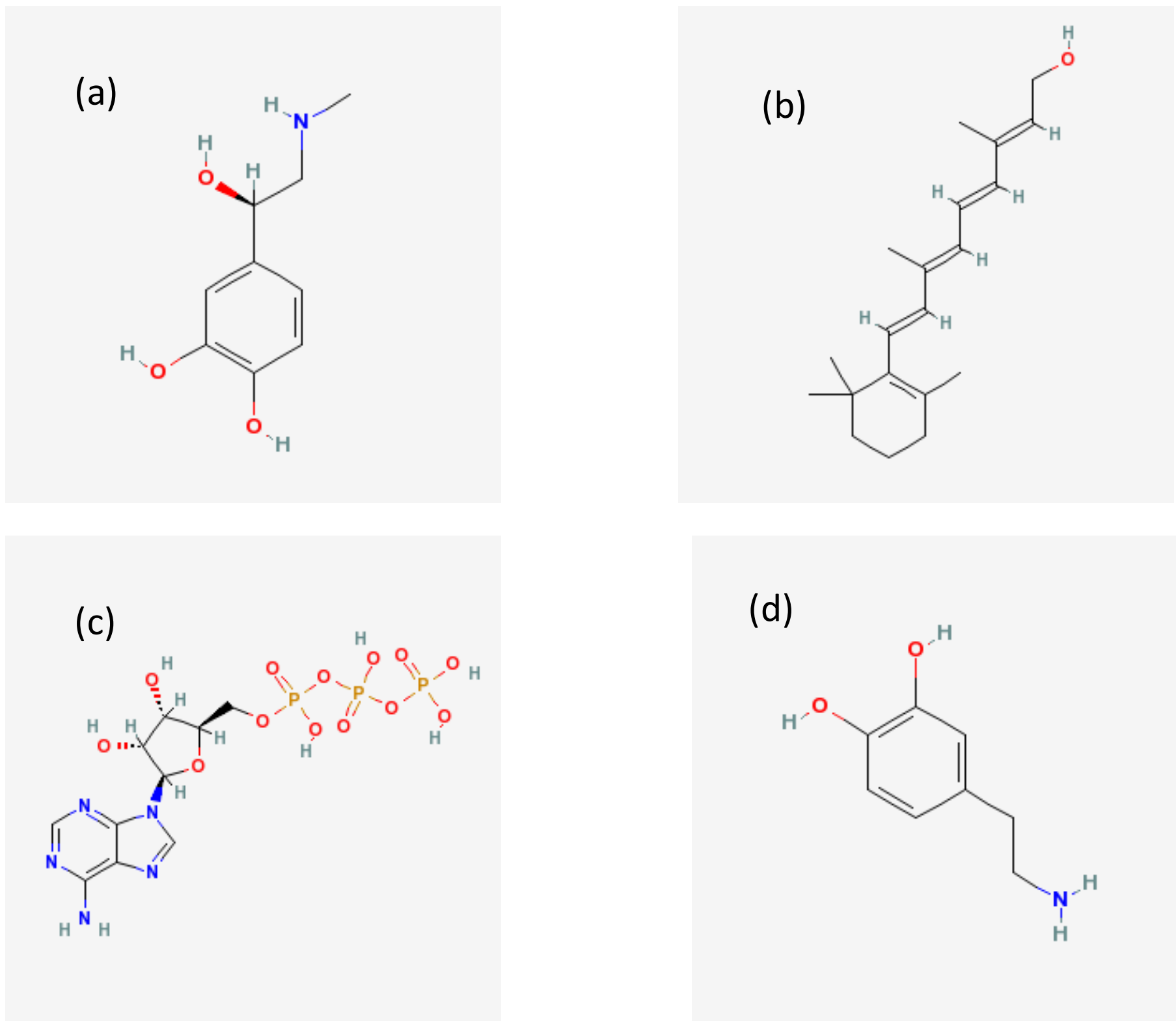

Fig. 3 (a) adrenalin (b) retinal (note the long aliphatic tail involved in optical absorption, and the absence of active amides) (c) adenosine triphosphate (ATP) (d) dopamine, which is similar to adrenalin.

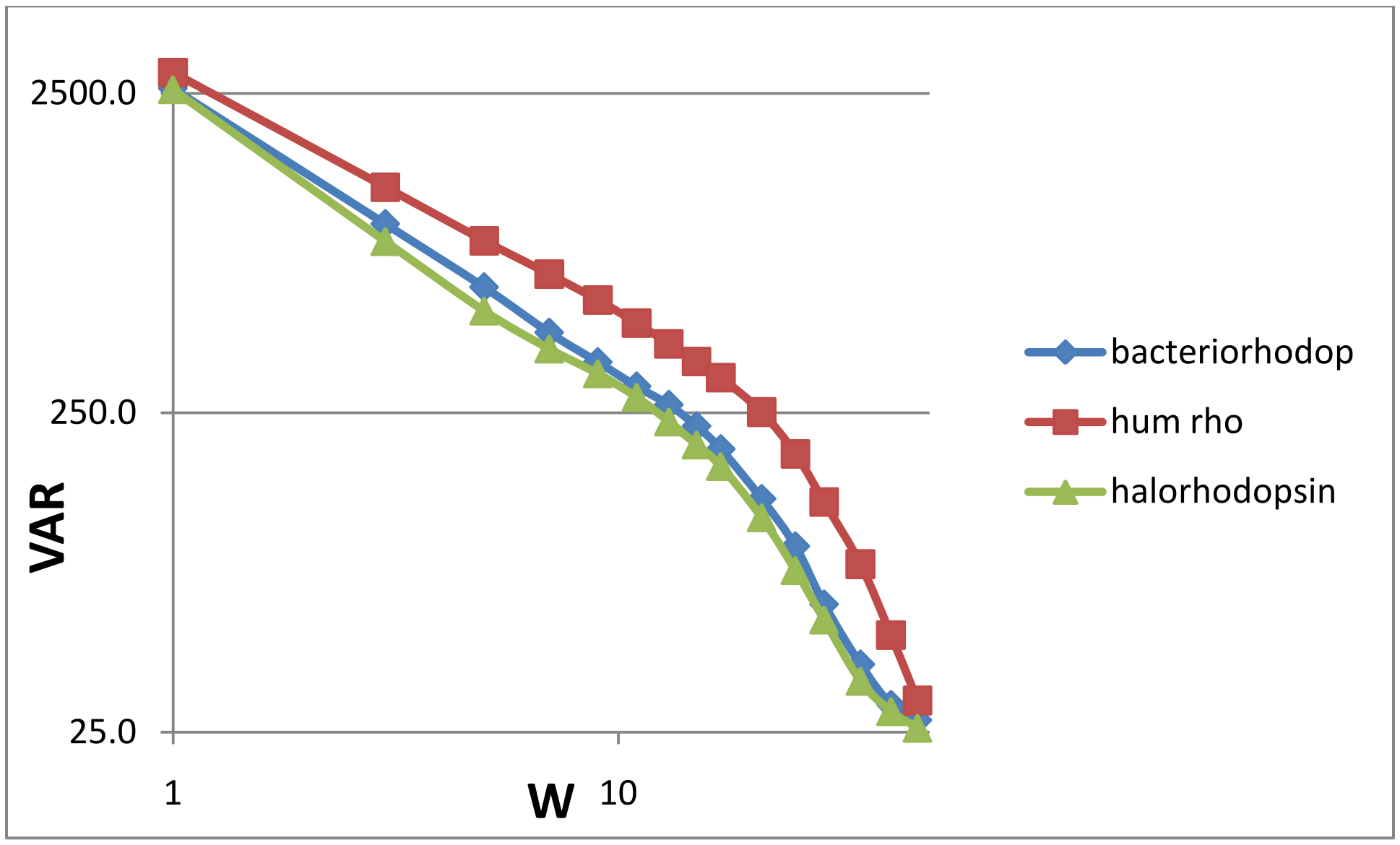

Fig. 4. Mammalian rhodopsins (here represented by human rhodopsin, P08100) exhibit slopes D**VAR**/D**W** slightly above -1 (in other words, hydropathically slightly ordered) for W ≤ 13, but for W > 13, the variances decrease rapidly (D**VAR**/D**W** ~ -2 to -3, large hydrosmoothing). The primitive rhodopsins show smaller hydrophobic fluctuations and a different structure, flattening out for W > 35. The primitive rhodopsins still incorporate retinal, but they do not couple on the cytoplasmic side to guanine $G_t$. The numbers of amino acids contained in the combined three cytoplasmic loops are Human rho, 53; bacteriorho, 43 and halorho, only 21. Correspondingly, although halorhodopsin is slightly larger than bacteriorhodopsin, it has the smallest VAR(47) of opsins (Table II). The smoothing of all rhodopsins above W = 13 is striking compared to other opsins. A fine point worth noting: this smoothing is probably connected to the larger size of retinal, especially the long aliphatic tail involved in optical transduction. Note that adenosine opsin shows a similar (but weaker) larger W smoothing, consistent with the large size of adenosine triphosphate.

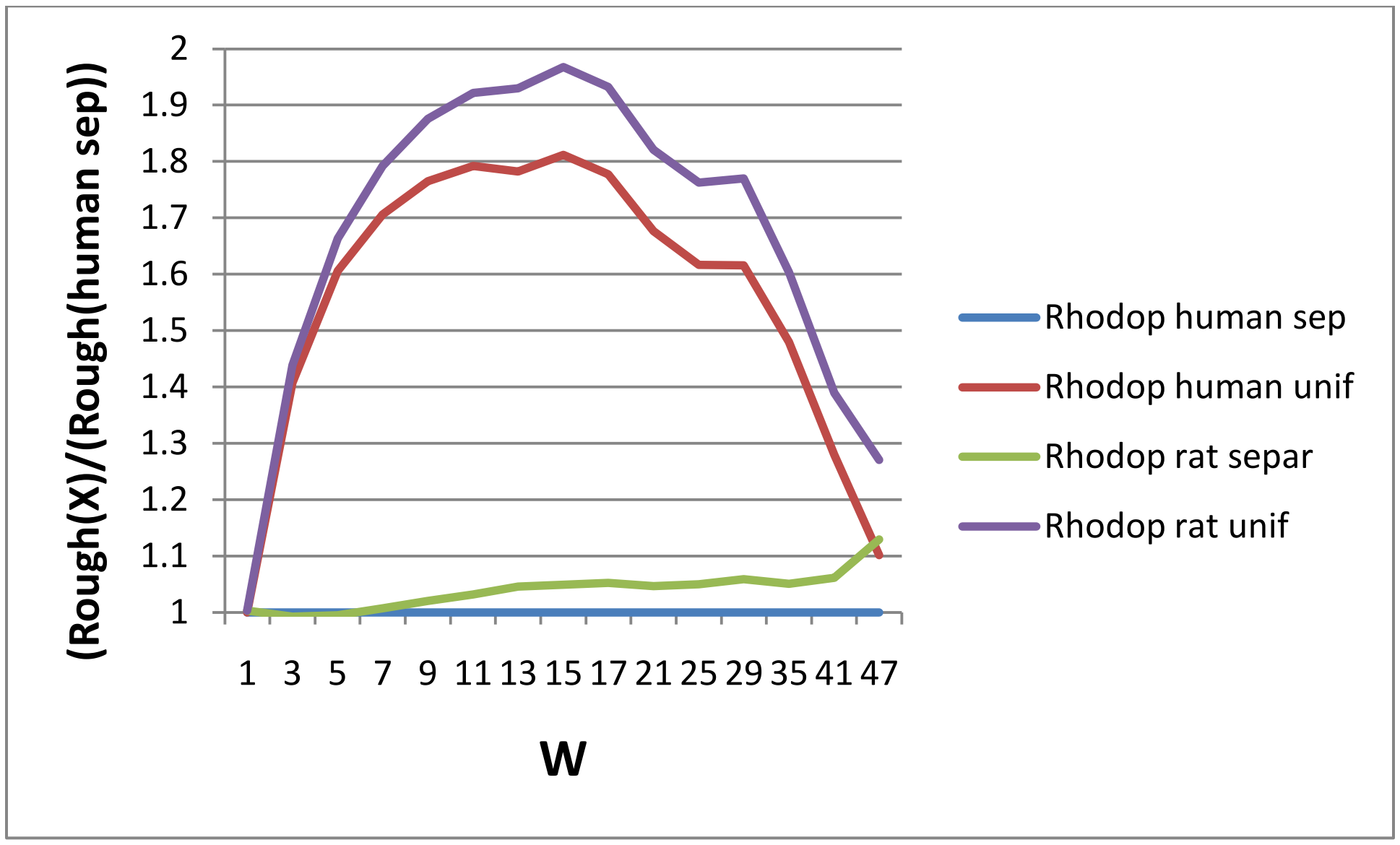

Fig. 5. The species differences between rat and human rhodopsin are small compared to the methodological differences between unified and separated definitions of roughness.

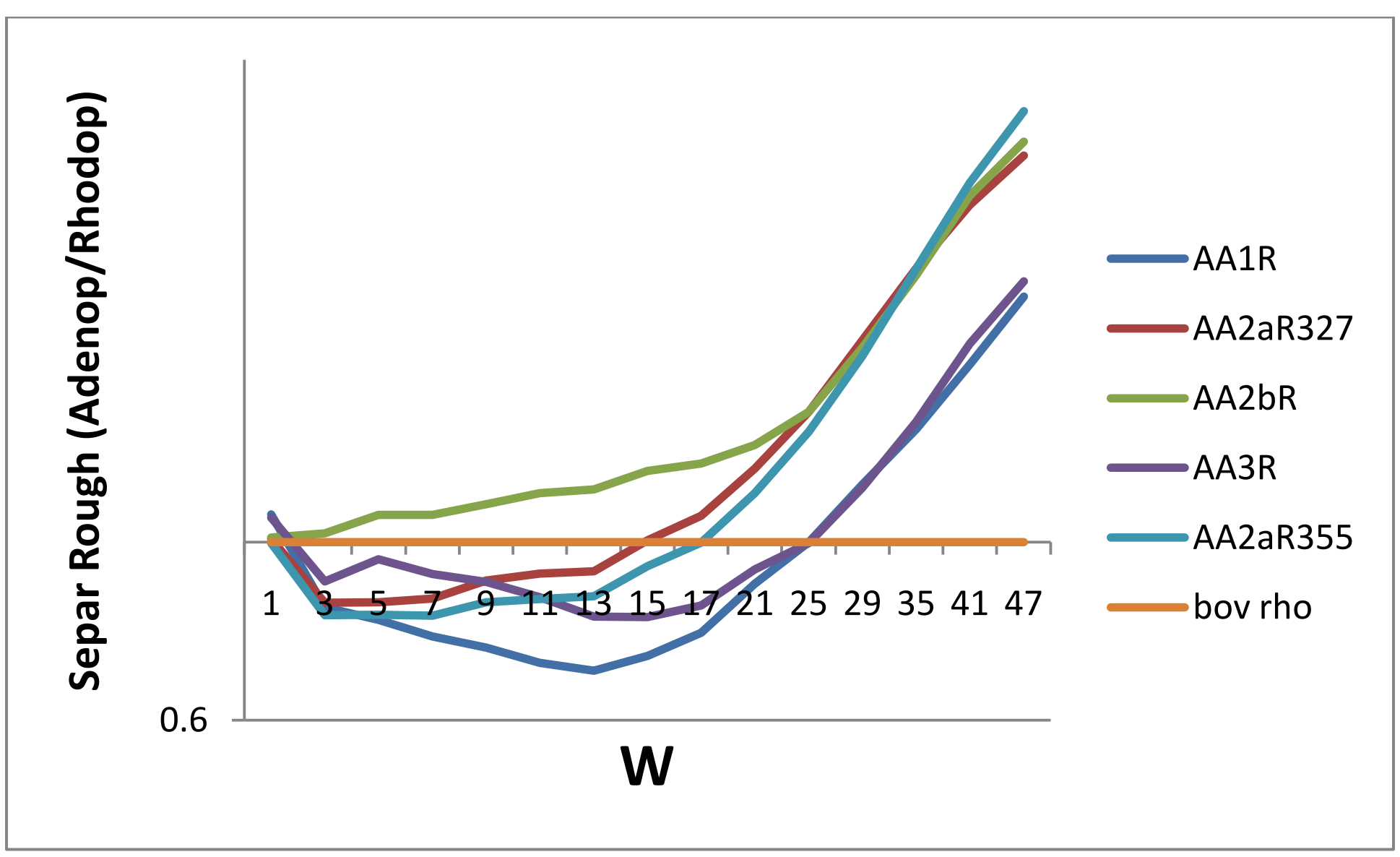

Fig. 6. Separated roughening profiles of human adenosine receptors normalized by bovine rhodopsin. Two cutoffs for the cytoplasmic terminal segment of AA2a are shown (327 and 355 residues), and one can see that the differences between them are small. For W < 7 there are two short-range groups, ([AA2b, AA3] and [AA1,AA2a]). The short- and long-range crossovers between AA2a and AA3 occur around W = 9. Beyond W = 15 the two groups ([AA1,AA3] and [AA2a,AA2b]) remain well separated, reflecting the hydrostability of loop-transmembrane interactions for length scales larger than transmembrane dimensions.

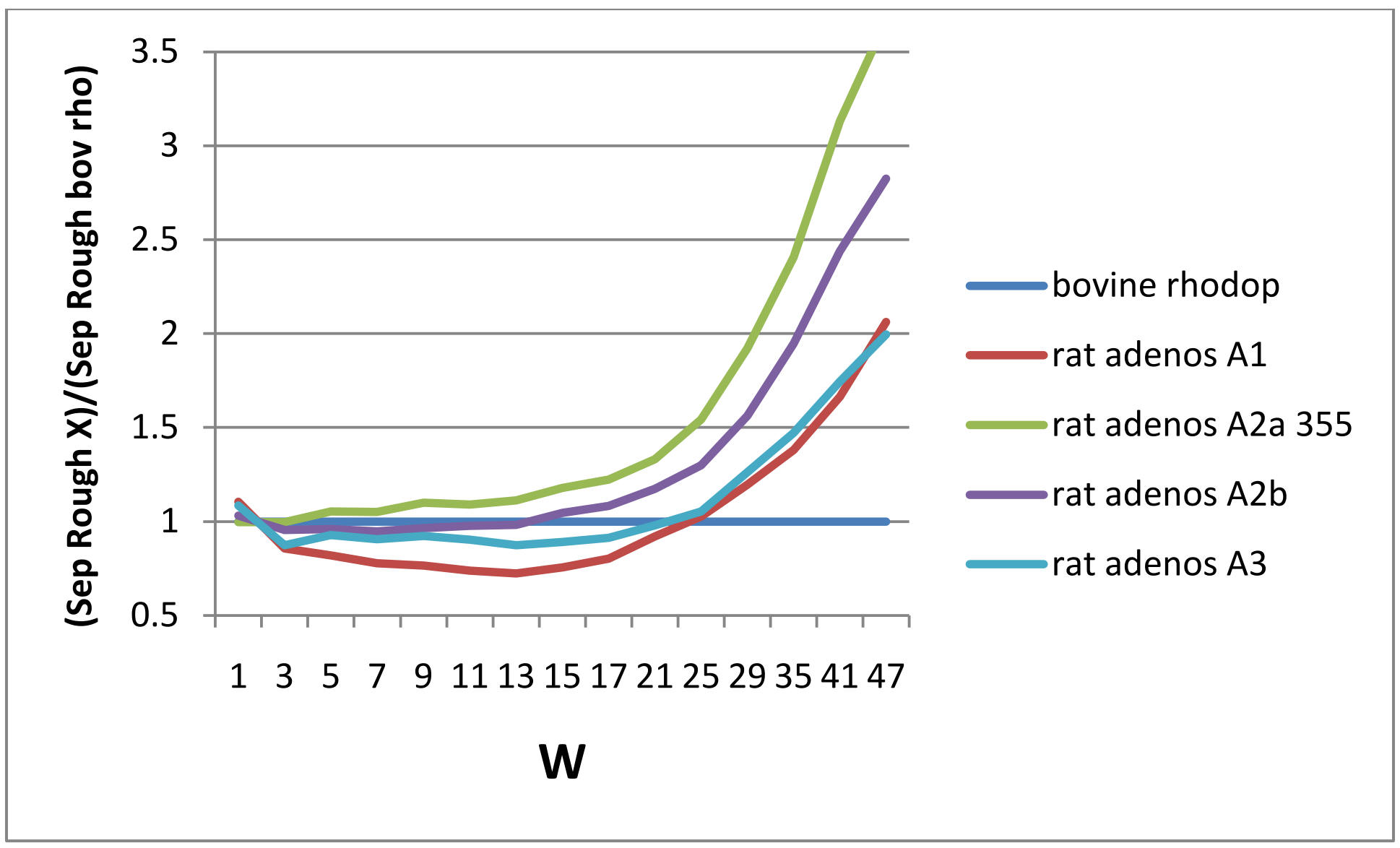

Fig. 7. Separated roughening profiles of rat adenosine receptors normalized by bovine rhodopsin. Comparison with Fig. 6 shows that the profiles for A1, A2b and A3 are little changed, but the rat profile for A2a is roughened relative to human A2a so much that the crossover noted in Fig. 6 has disappeared. The rat/human Blast sequence identities which are 82% for A2a and (94, 86, 74) for A1, A2b and A3, give no hint that A2a is special.

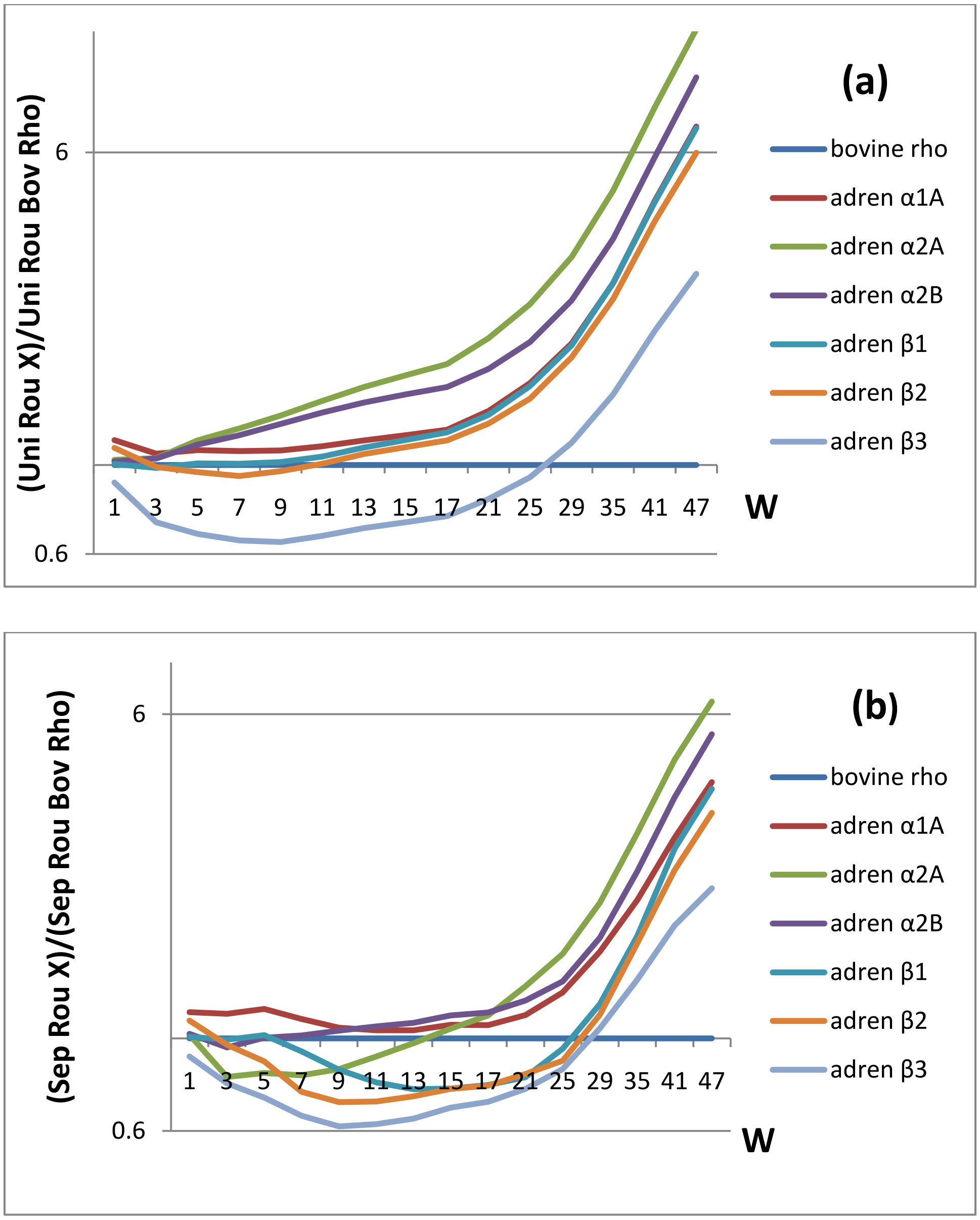

Fig. 8. Roughening profiles of human adrenergic receptors, normalized against bovine rhodopsin. In (a) the unified (variance) definition is used, whereas in (b) the proteins are separated into transmembrane, extracellular, and cytoskeleton segment groups for purposes of calculating three separate averages which are used to calculate average roughening as a function of window length W. Much more information (multiple crossings) about the differences within the adrenergic GPCR is exhibited in (b) than in (a). Near W = 20-25 (transmembrane lengths) in (a) the α and β subfamilies are poorly separated, while they are well separated in (b).

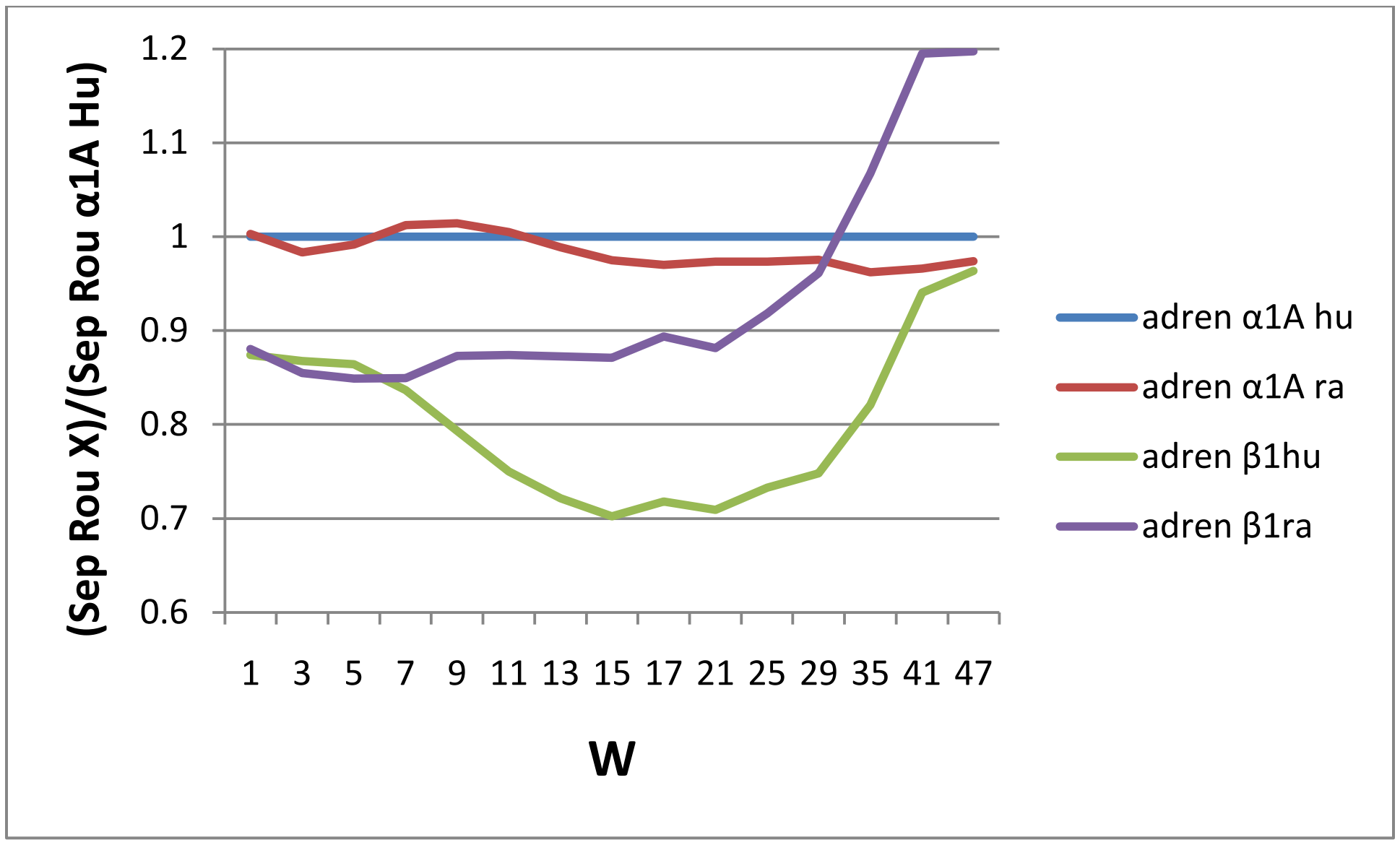

Fig. 9. SOC roughness plots of $\alpha_1$A and $\beta_1$ human and rat adrenergic families (normalized to $\alpha_1$A human).

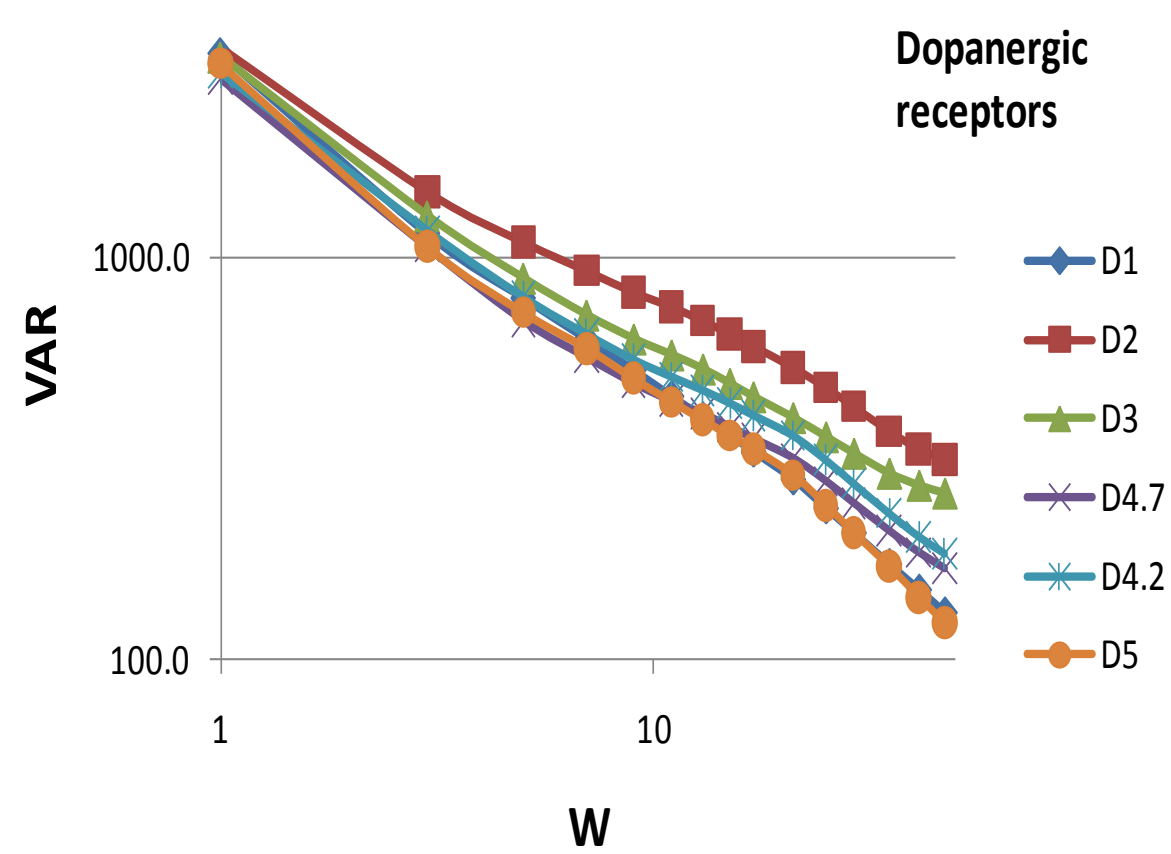

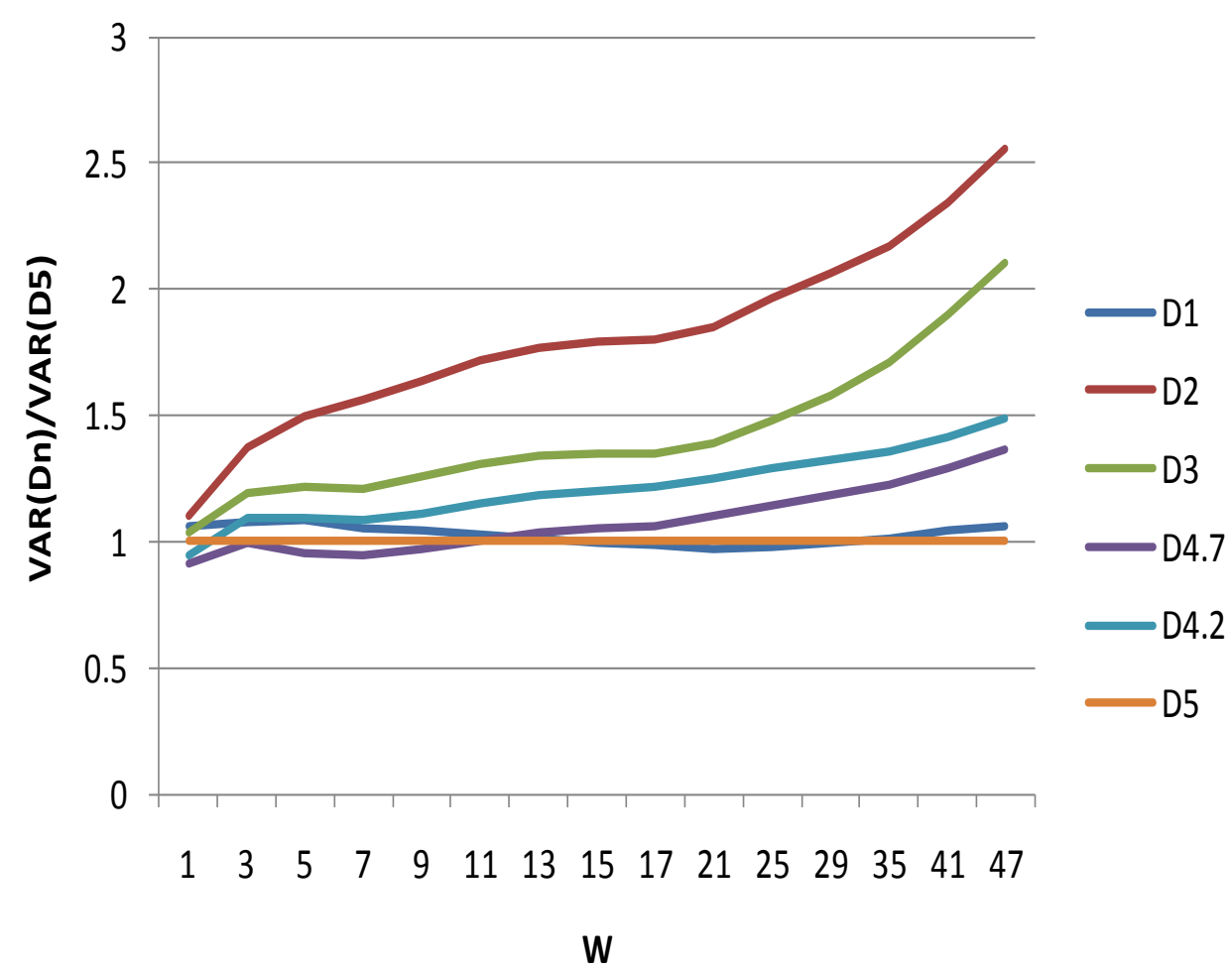

Fig. 10. Unified roughening profiles of human dopaminergic receptors. The differences between D4.2 and D4.7 are largest near W ~ 15, as expected for 16 aa tandem repeats.

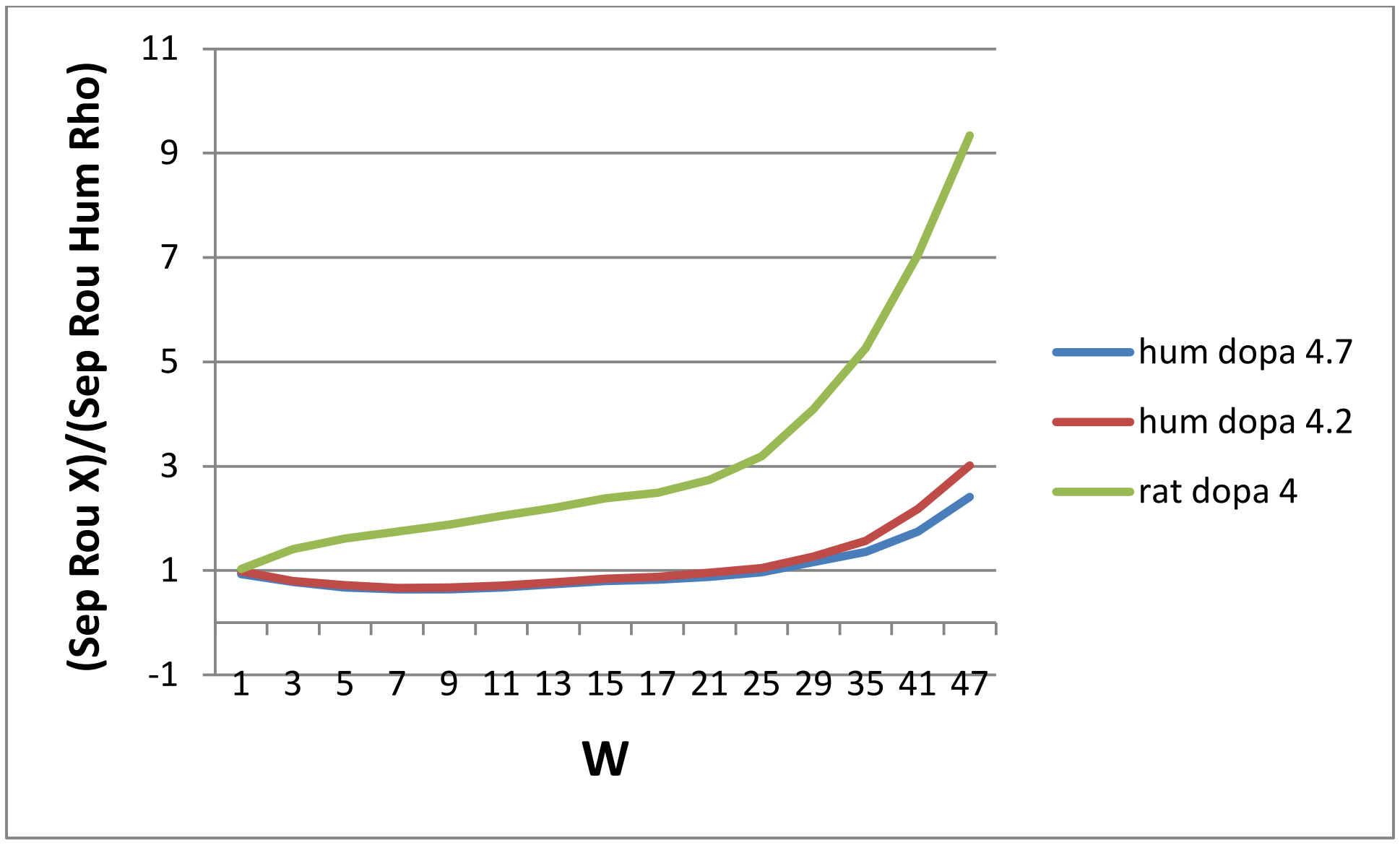

Fig. 11. The differences between human dopaminergic receptors D4.7 and D4.2 are small compared to their differences from rat D4.

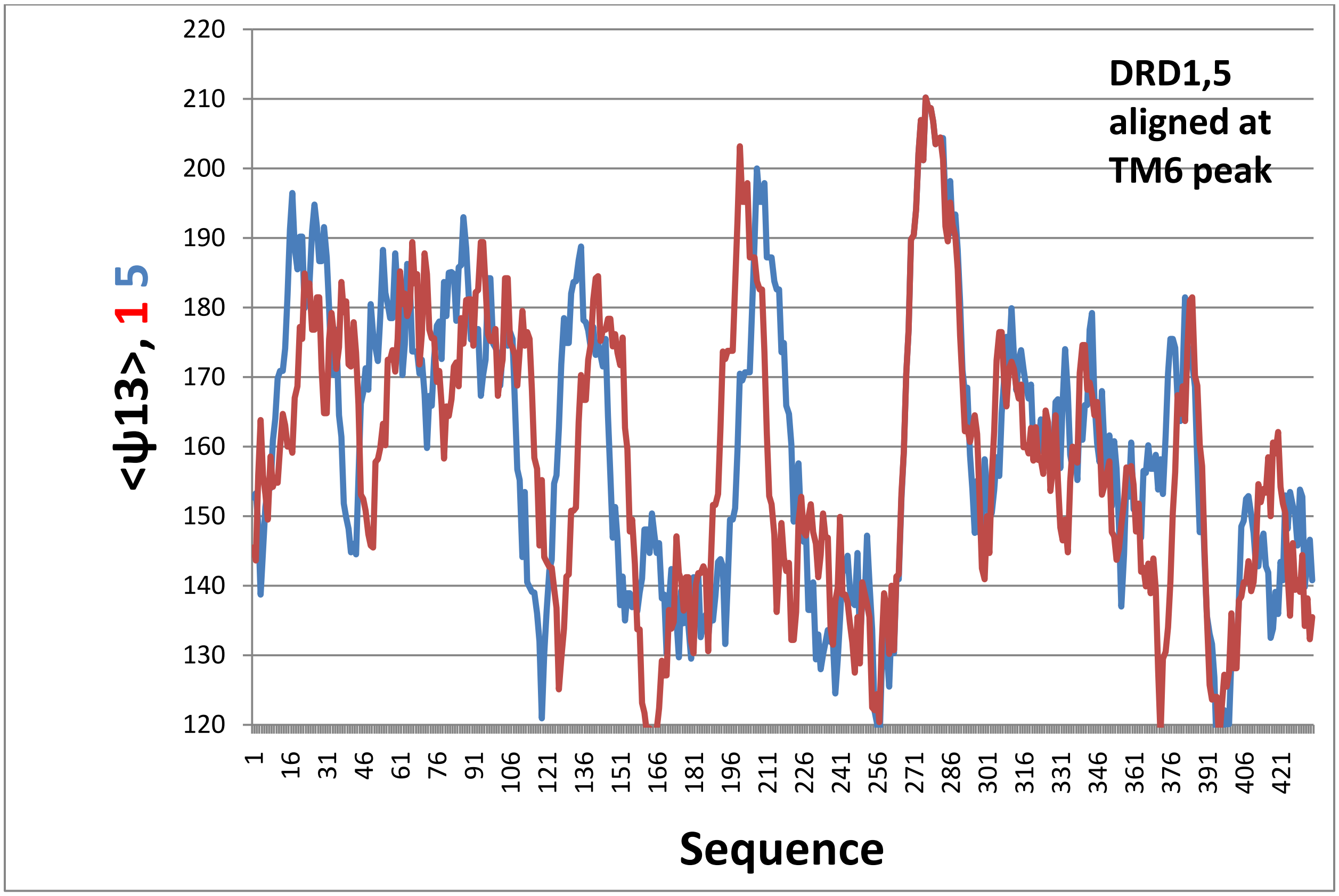

Fig. 12. Comparison of D1 and D5 < Ψ13> profiles.

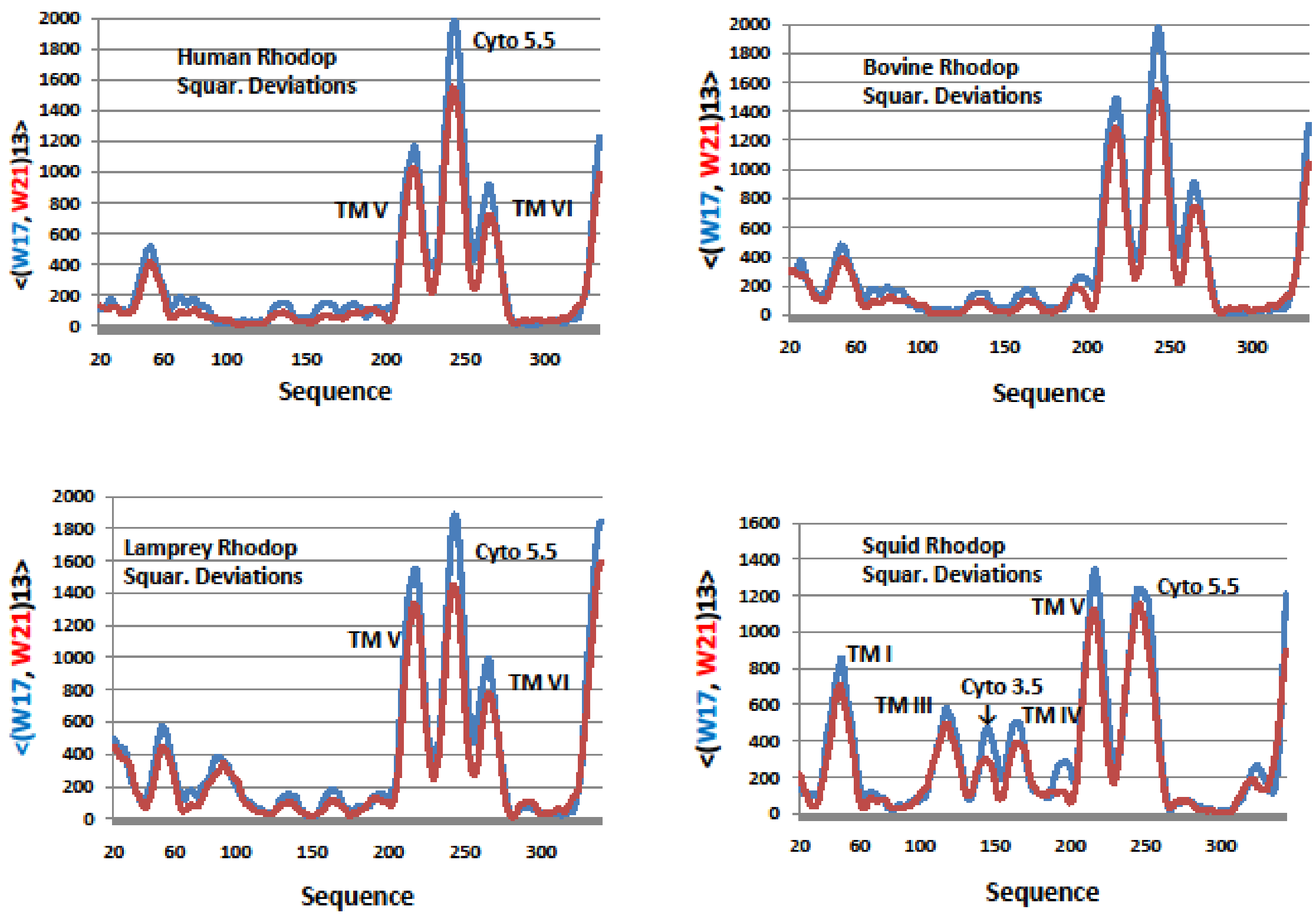

Fig. 13. Sequential distribution of squared deviations from mean for two windows, W = 17 and W = 21, smoothed over a third window $W_2$ = 13, for four rhodopsins. The vertebrate hydroprints are all similar and are dominated by TM V-VI structure, but the invertebrate squid rhodopsin hydroprint is qualitatively more widely distributed.

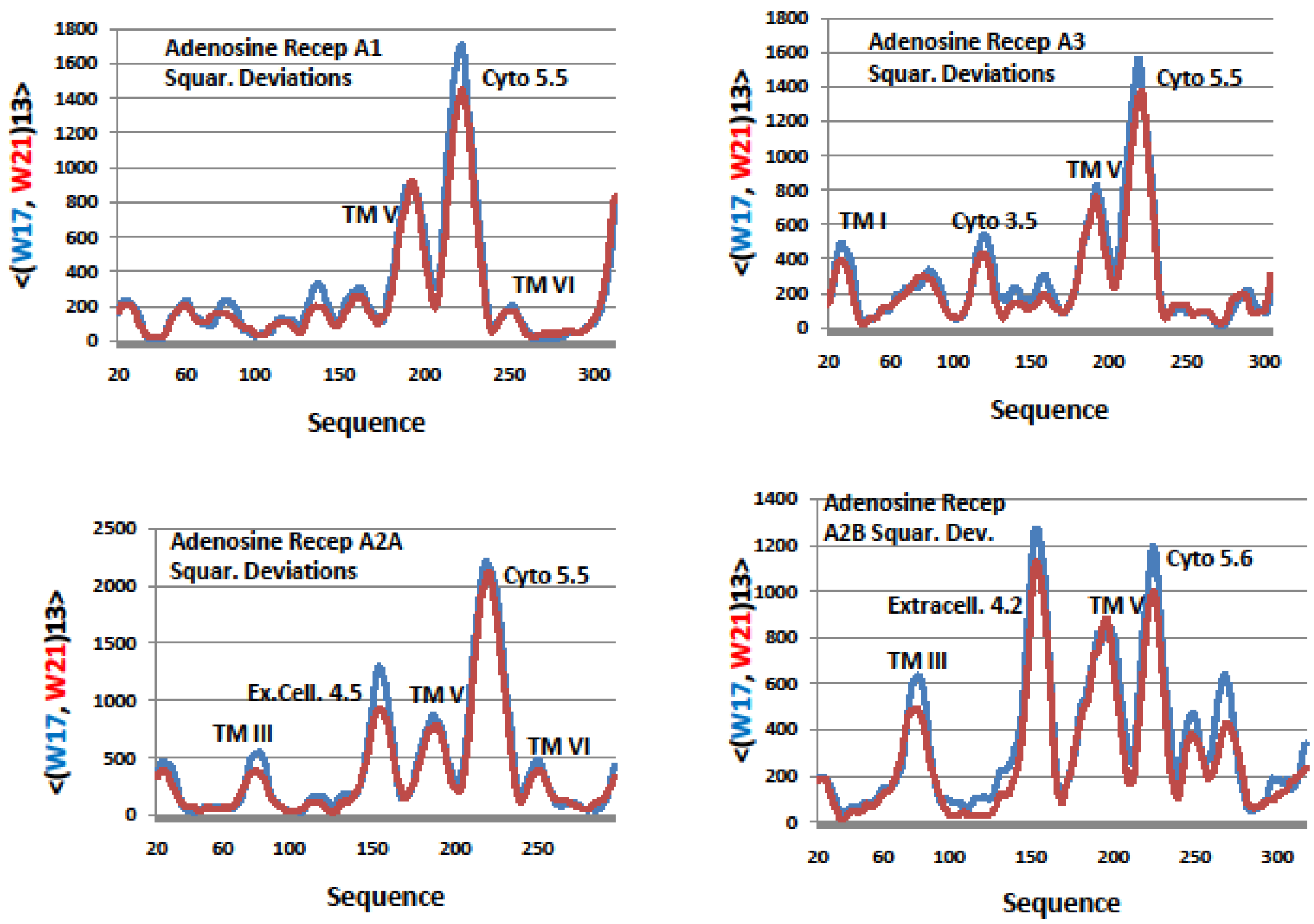

Fig. 14. The qualitative differences between the hydroprint profiles of adenosine receptors are consistent with stimulation (inhibition) of adenylate cyclase by [$A_{2A}$ and $A_{2B}$] ([$A_1$ and $A_3$]) receptors. The label "Extracell. 4.2" means that the peak occurs at 4.2, between the end of helix IV (4.0) and the beginning of helix V (5.0).

Table I. Here roughening variances are given for many rhodopsins, with unified (**separated**) definitions, and many evolutionary trends are obvious and expected, but some are surprising. There is a large difference at the in(vertebrate) squid(lamprey) transition, but lamprey still retains some squid roughness. The differences between cat and mouse are very small, much smaller than those between Atlantic and Pacific (colder water) Salmon, with colder water requiring a rougher profile. Note that fractional species differences are largest for W = 47, and these differences appear to be significant (not noisy), at least for rhodopsin. A few results for adenopsin and adrenopsin are also shown.

| Protein | Uniprot | Av. Hydro | W = 3 | W = 13 | W = 47 |
|---|---|---|---|---|---|
| Halorhodop | P16102 | 167.6 | 860 **711** | 235 **164** | 25.8 **24.9** |
| Bacteriorho | P02945 | 165.0 | 975 **728** | 265 **132** | 27.3 **18.4** |
| Squid Rho | P31356 | 163.4 | 1331 **815** | 482 **204** | 79.6 **83.2** |
| Lamp Rho | P22671 | 167.1 | 1320 **910** | 496 **276** | 53.5 **47.6** |
| Dolph Rho | O62791 | 167.4 | 1311 **931** | 440 **252** | 37.5 **32.9** |
| Whale Rho | O62793 | 167.8 | 1273 **905** | 413 **234** | 31.6 **27.9** |
| Atlan Salm | Q9IAH9 | 165.9 | 1219 **877** | 380 **201** | 43.1 **28.8** |
| Pacif Salm | Q6XR04 | 166.2 | 1221 **938** | 417 **256** | 48.4 **43.0** |
| Mouse Rho | P154409 | 167.4 | 1340 **939** | 458 **258** | 38.4 **35.0** |
| Cat Rho | Q95KU1 | 167.3 | 1332 **939** | 446 **251** | 37.3 **33.1** |
| Bovine Rho | P02699 | 167.6 | 1337 **926** | 452 **249** | 30.6 **27.9** |
| Human Rho | P08100 | 167.8 | 1270 **903** | 410 **230** | 31.4 **28.5** |
| Adeno A1h | P30542 | 166.4 | 1126 **768** | 360 **172** | 71.9 **56.5** |
| Aden A2ah | P29274 | 162.8 | 1160 **751** | 494 **213** | 163 **96.2** |
| Aden A2bh | P29275 | 167.7 | 1219 **950** | 427 **289** | 92.4 **88.2** |
| Ade A3h | P33765 | 177.6 | 1208 **828** | 378 **201** | 79.3 **59.0** |
| Adre α1Ah | P35348 | 159.5 | 1425 **1061** | 520 **260** | 213 **115*** |
| Adre α2Ah | P08913 | 153.5 | 1389 **749** | 707 **242** | 372 **180** |
| Adre α2Bh | P18089 | 153.8 | 1386 **881** | 646 **271** | 282 **150** |
| Adren β-1h | P08588 | 154.7 | 1313 **920** | 499 **188** | 211 **111** |
| Adren β-2h | P07550 | 159.8 | 1322 **894** | 481 **180** | 183 **97.1** |
| Adren β-3h | P13945 | 159.4 | 963 **722** | 315 **160** | 91.5 **64.0** |
| | | | | | |

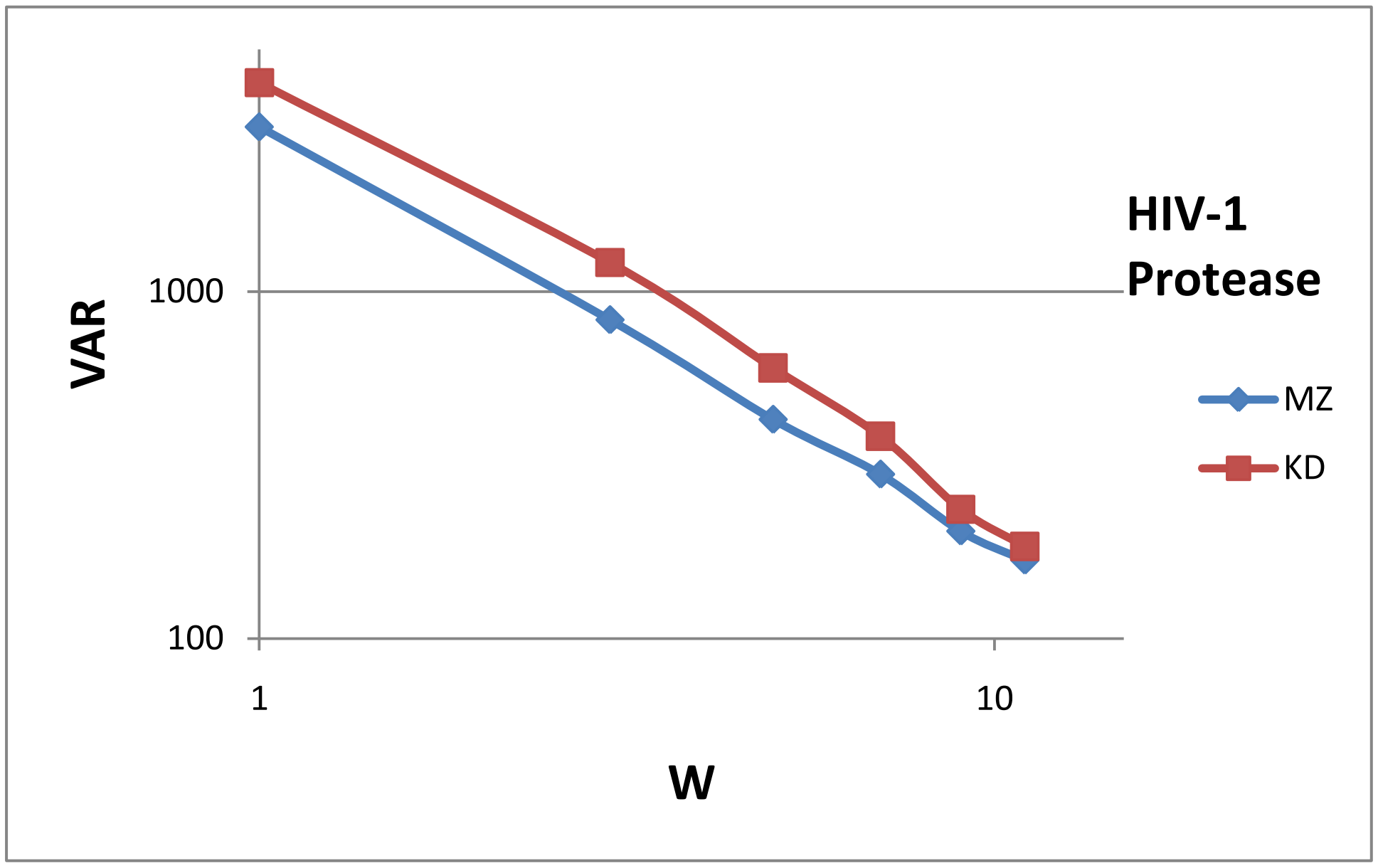

Fig. 9. Here we note a large difference between the two scales for 99-residue HIV-1B. Although 130 residue lysozyme *c* also shows a large difference between the two scales, the two differences behave much differently. In Fig. 6(a) for lysozyme *c* the two lines diverged beyond W = 9, whereas here they converge beyond W = 7. The most striking feature of the MZ line is the small bump at W = 7, which could be connected to incipient formation of α helices.

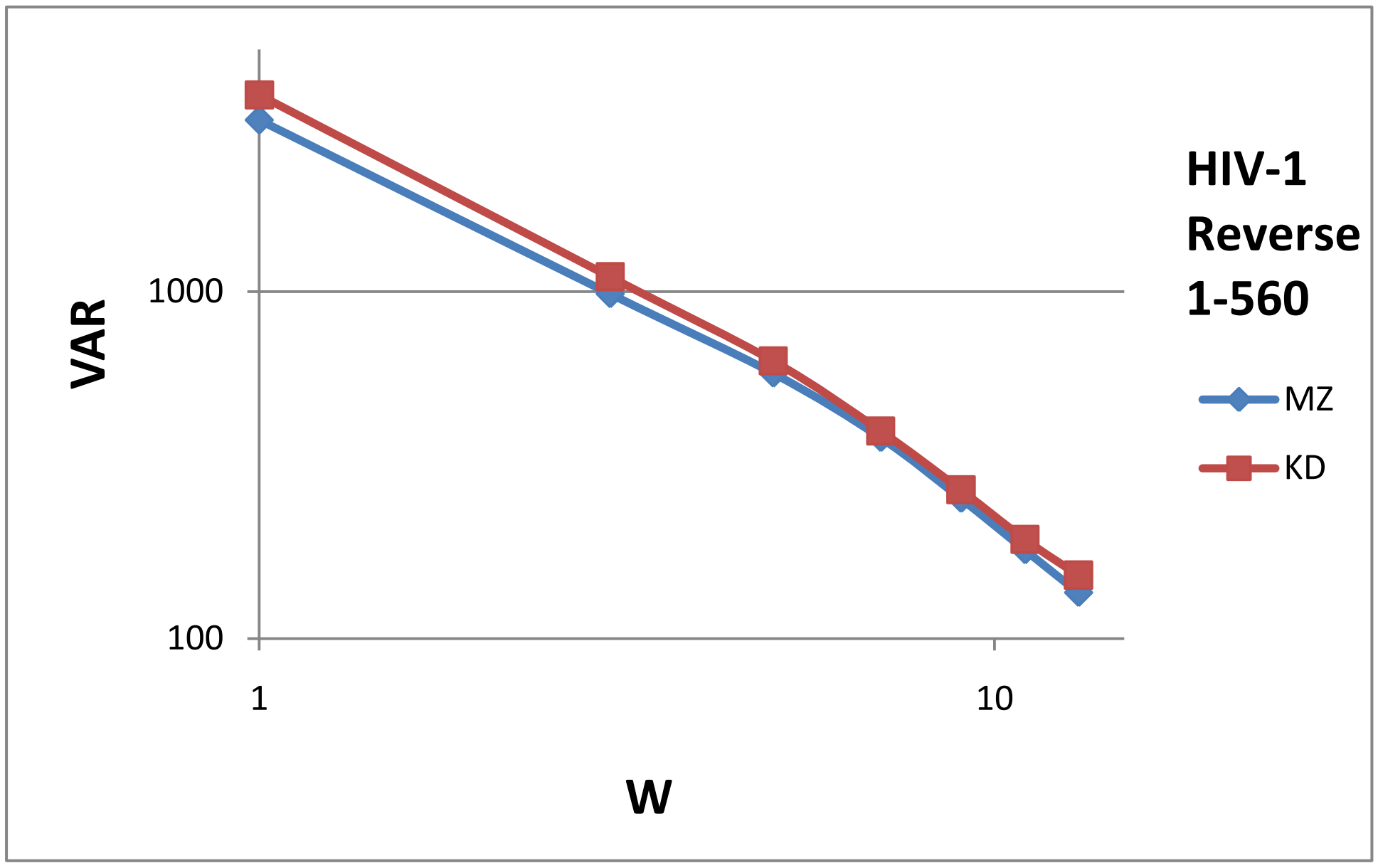

Fig. 10(a) Although the differences here are small, these curves exhibit two importtant features:the break in slope at W = 7, and the usual larger hydrosmoothing of the MZ scale compared to the KD scale.

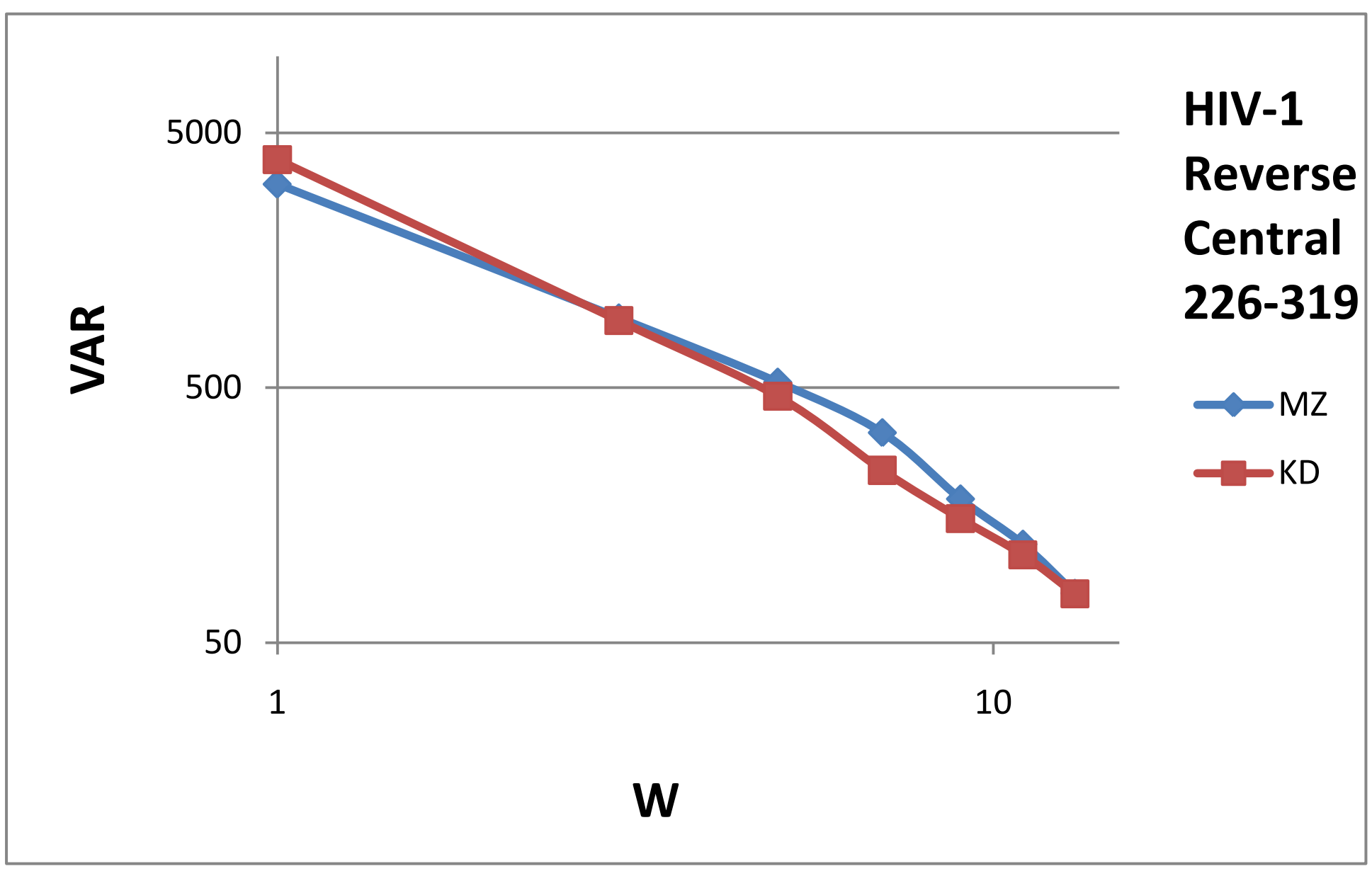

Fig. 10(b) Here something striking is shown, a break in slope of the KD curve at W= 5, which makes it drop below the MZ curve.

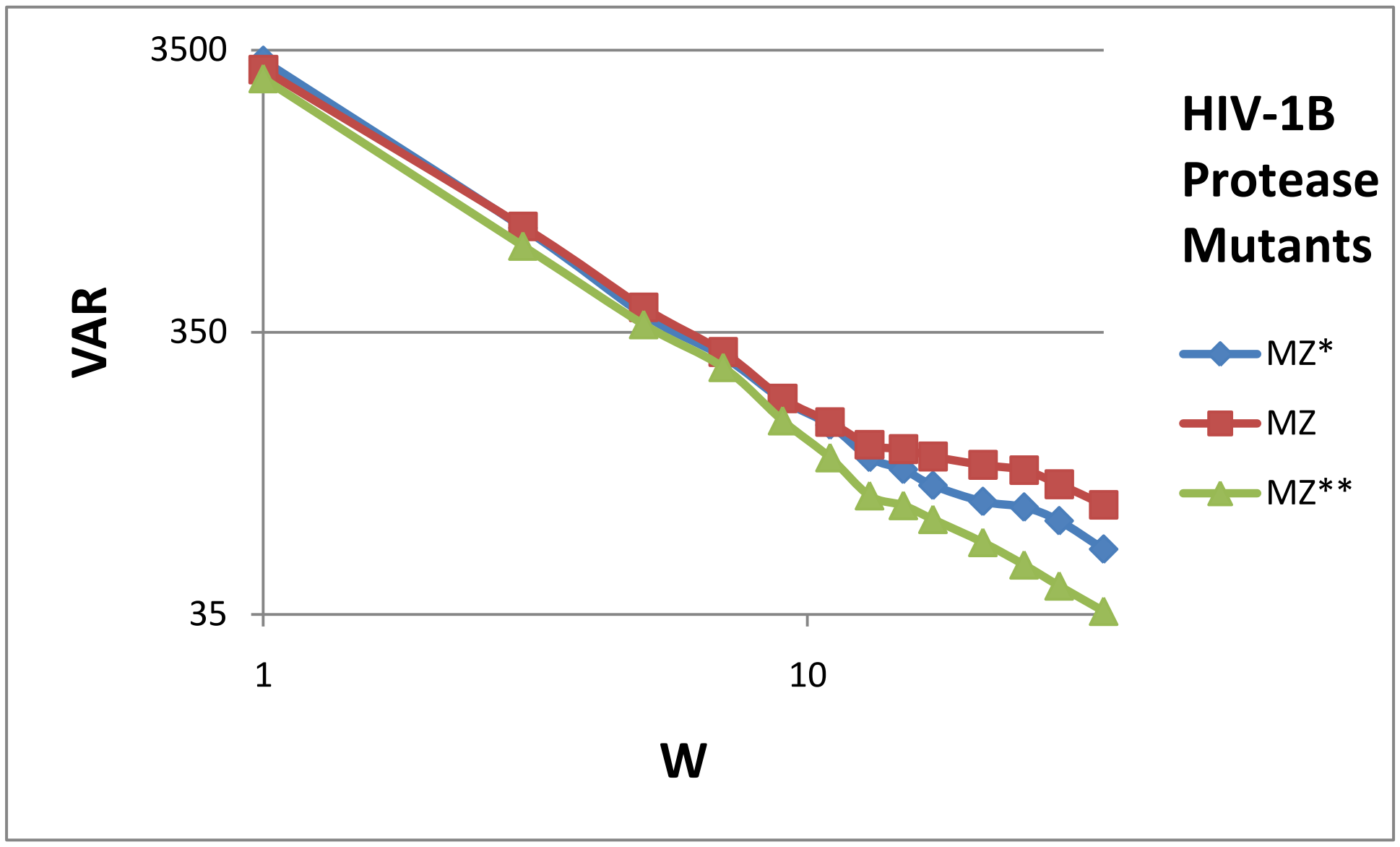

Fig. 11(a).

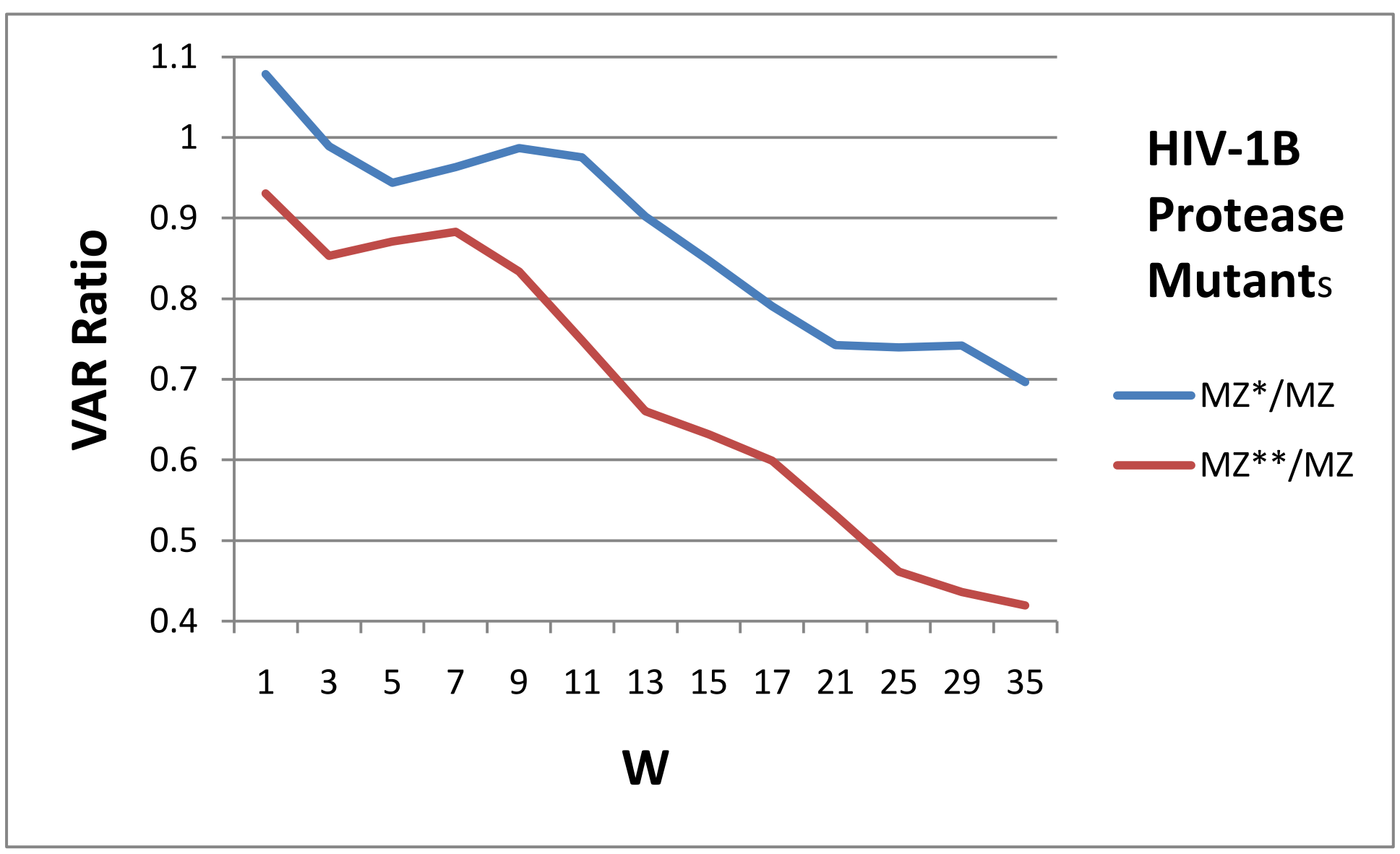

Fig. 11(b).

The carbazole ring system is oriented roughly perpendicular to the plane of the membrane, and the alkylamine chain (atoms 15 to 22 in the model) is nearly parallel to the heterocycle

Warne, T. *et al.* Structure of a $\beta_1$-adrenergic G-protein-coupled receptor. Nature 454, 486–491 (2008)

The $\beta_1$AR crystal structure shows the inactive state of the receptor, but it is notable that many agonists, including the natural ligands adrenaline and noradrenaline, are smaller than many of the best antagonists, including cyanopindolol. Agonists have a shorter distance, by two carbon–carbon bonds or 2–3 Å, between the catechol hydroxyl groups or their equivalent and the obligatory amine nitrogen.

Adren β-1 binds epinephrine and norepinephrine with approximately equal affinity.

The selectivity of beta-adrenoceptor antagonists at the human beta 1, beta 2 and beta 3 adrenoceptors

Author(s): Baker JG Source: BRITISH JOURNAL OF PHARMACOLOGY Volume: 144 Issue: 3 Pages: 317-322 Published: FEB 2005 Times Cited: 65 References: 27

Davis KL, Kahn RS, Ko G, Davidson M. Dopamine in schizophrenia: a review and reconceptualization. *Am J Psychiatry.* 1991;148(11):1474-1486.

Association of polymorphisms in the dopamine D4 receptor gene and the activity-impulsivity endophenotype in dogs Author(s): Hejjas K (Hejjas, K.), Vas J (Vas, J.), Topal J (Topal, J.), Szantai E (Szantai, E.), Ronai Z (Ronai, Z.), Szekely A (Szekely, A.), Kubinyi E (Kubinyi, E.), Horvath Z (Horvath, Z.), Sasvari-Szekely M (Sasvari-Szekely, M.), Miklosi A (Miklosi, A.) Source: ANIMAL GENETICS Volume: 38 Issue: 6 Pages: 629-633 Published: DEC 2007 Times Cited: 15 References: 21 Abstract: A variable number of tandem repeats (VNTR) polymorphism in exon 3 of the human dopamine D4 receptor gene (DRD4) has been associated with attention deficit hyperactivity disorder (ADHD). Rodents possess no analogous repeat sequence, whereas a similar tandem repeat polymorphism of the DRD4 gene was identified in dogs, horses and chimpanzees.